\documentclass[aps,amsmath,twocolumn,notitlepage,amssymb,prb,longbibliography]{revtex4-1}

\usepackage{graphicx}
\usepackage{dcolumn}
\usepackage{bm}
\usepackage[colorlinks=true,citecolor=red,urlcolor=blue]{hyperref}
\usepackage{wasysym}
\usepackage{bbold}
\usepackage{stmaryrd}
\usepackage{verbatim}
\usepackage{subfigure}

\begin{document}
\title{Emergent orbitals in the cluster Mott insulator on a breathing Kagome lattice}
\author{Gang Chen$^{1,2}$}
\author{Patrick A. Lee$^{3}$}
\affiliation{$^1$State Key Laboratory of Surface Physics, 
Department of Physics, Center for Field Theory \& Particle Physics,
Fudan University, Shanghai, 200433, China}
\affiliation{$^2$Collaborative Innovation Center of Advanced Microstructures, 
Nanjing, 210093, China}
\affiliation{$^3$Department of Physics, 
Massachusetts Institute of Technology,
Cambridge, Massachusetts, 02139, United States}

\begin{abstract}
Motivated by the recent developments on cluster Mott insulating  
materials such as the cluster magnet LiZn$_2$Mo$_3$O$_8$, 
we consider the strong plaquette charge ordered regime 
of the extended Hubbard model on a breathing Kagome lattice
and reveal the properties of the cluster Mottness. 
The plaquette charge order arises from the inter-site
charge interaction and the collective motion of three localized 
electrons on the hexagon plaquettes.
This model leads naturally to a reduction of the local moments by 2/3 
as observed in LiZn$_2$Mo$_3$O$_8$. Furthermore, at low temperatures 
each hexagon plaquette contains an extra 
orbital-like degree of freedom in addition to the remaining spin 1/2. 
We explore the consequence of this emergent orbital degree of freedom. 
We point out the interaction between the local moments is naturally 
described by a Kugel-Khomskii spin-orbital model. We develop a parton 
approach and suggest a spin liquid ground state with spinon Fermi 
surfaces for this model. We further predict an emergent orbital order when the system is under 
a strong magnetic field. Various experimental consequences for 
LiZn$_2$Mo$_3$O$_8$ are discussed,
including an argument that the charge ordering much be short ranged if 
the charge per Mo is slightly off stoichiometry.  
\end{abstract}

\date{\today}

\maketitle

\section{Introduction}

\begin{figure}[b]
{\includegraphics[width=6.5cm]{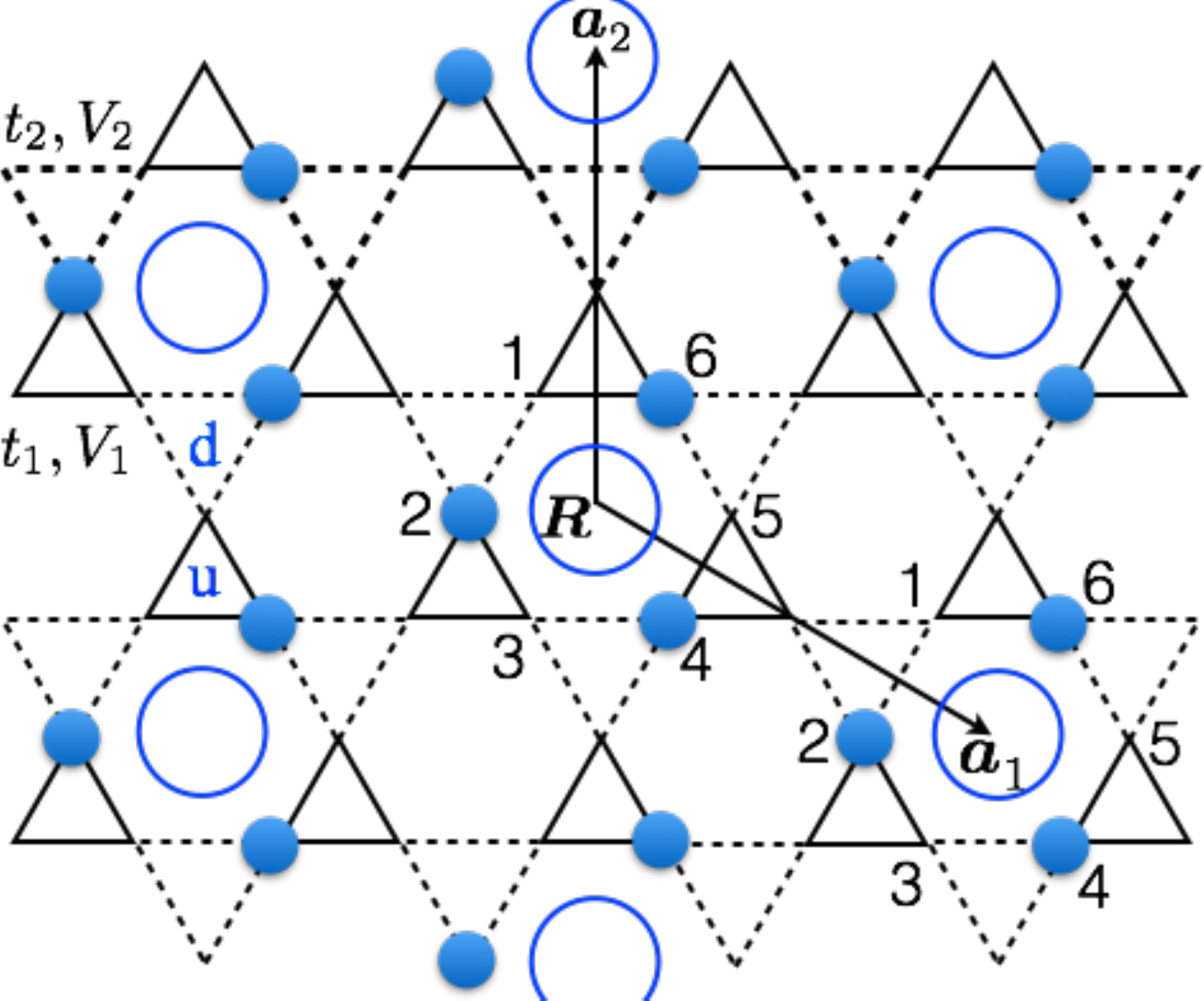}}
\caption{(Color online.) The breathing Kagome lattice
with plaquette charge order.  
The solid (dashed) lines represent the up (down) triangles.
The plauette charge order hosts three electrons that are resonating
on the hexagons with circles marked, and 
${\boldsymbol a}_1, {\boldsymbol a}_2$ are two lattice vectors 
that connect neighboring resonating hexagons. 
`${\boldsymbol R}$' labels the resonating hexagon, 
and `1,2,3,4,5,6' label the six vertices. 
}
\label{fig1}
\end{figure}

Spin, charge, and orbital are three basic degrees 
of freedom of condensed matter systems, and their mutual interaction, 
interplay, and entanglement cover the major topics of modern condensed 
matter physics~\cite{WCKB,PhysTMO,Nagaosa,Khaliullin}. 
In conventional Mott insulators, the electron charge 
localization creates the local spin moments at the lattice sites,
and the orbital degree of freedom becomes active when the local
crystal symmetry allows the degeneracy of the atomic orbitals~\cite{Nagaosa}.
Recently, the cluster Mott insulator emerge as a new
type of Mott insulator in which the electrons are localized 
inside the cluster~\cite{ChenPRL2014,ChenPRB2016,Lv2015,PhysRevB.96.054405,PhysRevLett.93.126403,
PhysRevLett.110.166402,PhysRevLett.93.126403,KimHS2014,PALeeLaw,CheongNature}. 
As a result, the keen interplay between the charge 
and the spin degrees of freedom in cluster Mott insulators (CMI) 
is often quite different from the conventional Mott 
insulator~\cite{ChenPRL2014,ChenPRB2016,Lv2015,PhysRevB.96.054405}.
In particular, it was shown that the two-dimensional CMI of 
the Kagome system~\cite{ChenPRB2016,PhysRevB.96.054405} 
with an extended Hubbard model at the $1/6$ electron filling may develop
 a plaquette charge order~\cite{Pollmann08,Isakov06,PhysRevB.90.035118,
PhysRevB.89.155141,PhysRevB.83.165118} on the hexagon plaquette (see Fig.~\ref{fig1}). 
This plaquette charge order immediately impacts the spin degree of freedom 
and modulates the spin properties by resconstructing the 
spin state within each plaquette. Such a charge-driven 
spin-state-reconstruction is one crucial property of the CMI in this 
system~\cite{ChenPRB2016}.  

Well known examples of the cluster magnets include 
LiZn$_2$Mo$_3$O$_8$, Li$_2$InMo$_3$O$_8$~\cite{GALL201399},
and ScZnMo$_3$O$_8$~\cite{doi:10.1021/ic00198a009}, 
where the Mo electrons are in the CMI with the Mo electrons 
localized in the smaller triangular clusters of the distorted Kagome lattice 
(see Fig.~\ref{fig1})~\cite{Mourigal14,Sheckelton14,Sheckelton12,Mcqueen2015,Quilliam}.
The distortion is such that the up and down triangles have different bond lengths and the lattice is often referred to as the breathing kagome. 
Interestingly, the material LiZn$_2$Mo$_3$O$_8$ experiences two Curie regimes with distinct
Curie-Weiss temperatures and Curie constants~\cite{Sheckelton14,Sheckelton12}
in which the low temperature Curie constant is 1/3 of the high temperature one
and the low temperature Curie-Weiss temperature is much smaller than 
the high temperature one. Moreover, the system remains magnetically 
disordered down to the lowest measured temperature, and the inelastic 
neutron scattering does observe a continuum of excitations~\cite{Mourigal14}. 
This is consistent with the proposal of spin liquid ground state in this material. 
Partly inspired by the experiments in LiZn$_2$Mo$_3$O$_8$,
we here explore the strong plaquette charge ordered 
regime of the CMI on the breathing Kagome system 
where the electron charges are localized 
on the resonating hexagon plaquettes (see Fig.~\ref{fig1}).   
In addition to the on-site repulsion, a large inter-site 
repulsion is assumed which forbids the occupation of neighboring sites. 
This leads to plaquette charge ordering and the expansion of the unit cell, 
formed by a triangular lattice of hexagons marked by the circles in Fig~\ref{fig1}. 
The low lying degree of freedom is the collective resonant rotation 
of the three occupied sites on each hexagon (see Fig.~\ref{fig2}). 
To put this model in context of the earlier model by Flint and Lee~\citep{Flint13}, 
there the inter-site repulsion is assumed to be weak and each up
 triangle is occupied by one electron, and no correlation is assumed 
 around the hexagons. The up triangles form a triangular lattice and
 a lattice distortion is postulated which creates a honeycomb lattice 
 of up triangles, with the spin at the center of the honeycomb relatively 
 isolated and responsible for the local moments at low temperatures. 
 Note that both for this model and the current model, a tripling of 
 the unit cell is assumed. This has been searched for by X-ray scattering 
 but so far no new diffraction peaks have been observed. 
 This issue will be discussed in the Discussion section, 
 where we point out that if the system is slightly off stoichiometry, 
 domain  walls will form between the ordered states. 
Due to a special feature of domain walls forming a honeycomb lattice~\cite{PhysRevB.25.349}, 
it can be shown that long range order is always destroyed, {\sl i.e.} 
the system can only have short range order. This may help explain 
the absence of new diffraction spots and both models may remain viable. 
We also point out that in the Flint-Lee model addressed only the freeze 
out of 2/3 of the spins at low temperatures, and the ultimate fate of the  
local moments that remained was not discussed. In the current model, 
we address both the freeze out and the true ground state of this system 
and argue that due to an emergent orbital degree of freedom, a spin liquid 
state may form as the true ground state.

We also compare the current paper with a previous work on a similar model~\cite{ChenPRB2016}
which treats the weak plaquette order regime.
The current treatment of the CMI is analogous to the strong Mott 
regime of a conventional Mott insulator, while the previous 
weak plaquette charge ordered regime~\cite{ChenPRB2016} 
is like the weak Mott regime (ie close to the Mott transition)
where the charge fluctuation may destabilize the spin order and 
lead to a spin liquid~\cite{Motrunich05,Lee05}. 
We find that in the  strong charge ordered regime, the charge-spin interaction appears in a 
much more straightforward and transparent manner. We explain the local moment 
reconstruction in the presence of the strong plaquette charge order on the hexagon, giving rise to
 a net spin-1/2 local moment on the hexagon. We point out 
that there exists an emergent orbital-like degree of freedom. This emergent orbitals 
are two-fold degenerate and protected by the symmetry of the hexagon plaquette. 
The natural model, that describes the interaction between the effective spin and 
the emergent orbital on the hexagon plaquette, is the Kugel-Khomskii exchange model~\cite{Kugel82}. 
As a comparison with the conventional Mott insulators, 
the Kugel-Khomskii model is used to describe the exchange interaction
between the local moments when there exists an
orbital degeneracy for the atomic orbitals~\cite{Kugel82}. 

For the Kugel-Khomskii model, we design a fermionic parton approach 
to represent the effective spin and the emergent orbital degrees of 
freedom, and propose a spinon Fermi surface spin liquid ground state. 
We point out that the emergent orbital generically creates 
non-degenerate spinon bands and allows inter-band particle-hole
excitations. Specifically, the inter-band particle-hole
excitations would manifest as a finite-energy spinon continuum 
at the $\Gamma$ point in the inelastic neutron scattering and 
the optical measurement. Polarizing the spin degrees of freedom by applying strong 
magnetic fields, we obtain a simple 120-degree compass model 
for the emergent orbital interaction. We further predict that 
the system selects a specific orbital order via order by quantum 
disorder and supports a nearly gapless pseudo-Goldstone mode. These 
results establish a new perspective on the Mottness of the CMI. 

The following part of the paper is organized as follows. In Sec.~\ref{sec2},
we introduce the extended Hubbard model and explain the plaquette charge order.
In Sec.~\ref{sec3}, the explain the local moment structure of the resonating hexagon
in the strong plaquette charge ordered regime and point out the 
fundamental existence of the emergent orbital degree of freedom. 
In Sec.~\ref{sec4}, we derive the Kugel-Khomskii model that describes the 
exchange interaction between the spin and the orbital on the triangular lattice
formed by the resonating hexagons. In Sec.~\ref{sec5}, we design a 
parton construction and suggest the features of the spinon continuum 
for the proposed spinon Fermi surface ground state. In Sec.~\ref{sec6}, 
we explain the emergent orbital order, quantum order by disorder
effect of the compass model for the orbitals, and the orbital 
excitation when the spin is polarized by the external magnetic field. 
In Sec.~\ref{sec7}, we discuss the {the relevance of this model to 
LiZn$_2$Mo$_3$O$_8$ and explore various experimental consequences.
We end with a a broad view on the cluster Mott insulating materials.

\section{The microscopic model and the plaquette charge order}
\label{sec2}

We start with the extended Hubbard model
on the breathing Kagome lattice (see Fig.~\ref{fig1}), 
\begin{eqnarray}
H &=&  -\sum_{ \langle ij \rangle \in \text{u}} 
(t_1^{} c^{\dagger}_{i\sigma} c^{\phantom\dagger}_{j\sigma} + h.c.)
- \sum_{ \langle ij \rangle \in \text{d}} 
(t_2^{} c^{\dagger}_{i\sigma} c^{\phantom\dagger}_{j\sigma} + h.c.)
\nonumber \\
&& 
+ \sum_{\langle ij \rangle \in \text{u}} V_1^{} n_i n_j 
+ \sum_{\langle ij \rangle \in \text{d}} V_2^{} n_i n_j  
+ \sum_{i} U n_{i\uparrow}n_{i\downarrow} ,
\label{eq1}
\end{eqnarray}
where $c^{\dagger}_{i\sigma}$ ($c^{\phantom\dagger}_{i\sigma} $) 
creates (annihilates) an electron with spin $\sigma$($=\uparrow,\downarrow$) 
at the lattice site $i$, $n_i$($\equiv n_{i\uparrow} + n_{i\downarrow}$) 
is the electron occupation number, and `u' and `d' refer to the up and down 
triangles that are of different sizes, respectively.
Here, $t_1$ and $V_1$ ($t_2$ and $V_2$) are the electron hopping and repulsion 
on neighboring sites of the up (down) triangles, respectively. The electron 
filling is 1/6, {\it i.e.} one electron per unit cell on the breathing Kagome lattice. 
This model was suggested to capture the physics of the Mo-based cluster magnets 
such as LiZn$_2$Mo$_3$O$_8$ in which the Mo atoms form an breathing Kagome 
lattice~\cite{ChenPRB2016,GALL201399,doi:10.1021/ic00198a009}.  

\begin{figure}[t]
{\includegraphics[width=6.5cm]{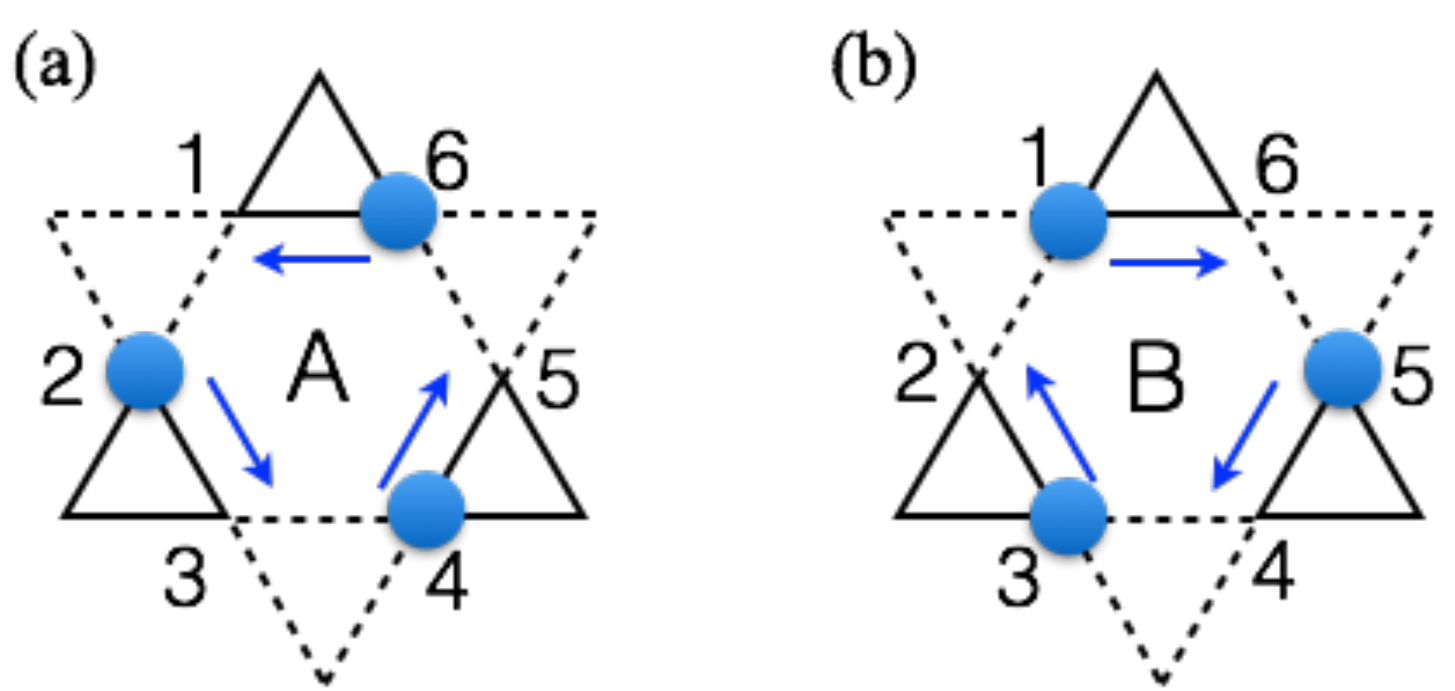}}
\caption{(Color online.) The correlated and collective motion of the 
three electrons on the elementary hexagon. Arrow indicates the hopping 
direction. Note that the hoppings of the three electrons happen at the 
same time. 
}
\label{fig2}
\end{figure}

The Hubbard $U$ interaction for our system merely removes the electron double 
occupancy on the lattice site, but cannot localize the electrons on the lattice sites. 
The electrons can move on the lattice without encountering 
any double occupancy. This is quite different from the conventional Mott insulator 
where the electrons are localized on the lattice sites. 
It is the inter-site interactions, $V_1$ and $V_2$, 
that localize the electron on the triangular clusters of the Kagome system. 
Despite being localized on the triangular clusters, 
the electrons manage to fluctuate in a collective fashion due to 
the extensive degeneracy of the electron occupation configuration on the 
Kagome lattice. As $U$ is often quite large compared to $t_1,t_2,V_1,V_2$,
one could safely ignore the electron configurations with any double occupancy. 
With a third-order degenerate perturbation of the electron hoppings, 
we obtain an effective Hamiltonian that operates on the degenerate 
electron occupation manifold and is given as~\cite{ChenPRB2016}
\begin{eqnarray}
H_{\text{eff}} &=& 
- \sum_{\hexagon} \sum_{\alpha \beta \gamma }
\big[
{K}_1  (c^{\dagger}_{1\alpha} c^{\phantom\dagger}_{6\alpha} 
                                 c^{\dagger}_{5\beta} c^{\phantom\dagger}_{4\beta}  
  c^{\dagger}_{3\gamma} c^{\phantom\dagger}_{2\gamma} + h.c.) 
\nonumber \\
&& \quad\quad\quad \,\,\,\, +  
{K}_2  (c^{\dagger}_{1\alpha} c^{\phantom\dagger}_{2\alpha} 
                                 c^{\dagger}_{3\beta} c^{\phantom\dagger}_{4\beta}  
                                  c^{\dagger}_{5\gamma} c^{\phantom\dagger}_{6\gamma}  
+ h.c.) 
\big],
\label{eq2}
\end{eqnarray}
where we have
\begin{eqnarray}
{K}_1 = {6t_1^3}/{V_2^2}, 
\quad\quad  
{K}_2 = {6t_2^3}/{V_1^2},
\end{eqnarray} 
and  ``1,2,3,4,5,6'' refer to the six vertices on the elementary hexagon of the Kagome lattice. 
$H_{\text{eff}}$ describes the correlated and collective motion of the three electrons
on the elementary hexagon (see Fig.~\ref{fig2}). 
By mapping the electron occupation to the dimer covering on the dual honeycomb 
lattice~\cite{ChenPRB2016,PhysRevB.64.144416},
the previous work has obtained a plaquette charge order where the electrons 
preferentially occupy 1/3 of the hexagons in a periodic fashion 
(see Fig.~\ref{fig1})~\cite{ChenPRB2016,PhysRevB.89.155141,PhysRevB.90.035118,
PhysRevLett.100.136404,PhysRevB.83.165118}. This plaquette charge order is 
a quantum effect because the three electrons are resonating on the hexagons 
and form a linear superposition of the two occupation configurations~\cite{ChenPRB2016}. 
In the strong plaquette charge ordered limit, the electron (charge) occupation 
wavefunction would be well approximated by a simple product state with 
\begin{eqnarray}
{| \Psi \rangle_{\text c}^{} } = \prod_{\boldsymbol{R}} 
\frac{1}{\sqrt{2}} \big[  
                      | \hexagon_{\boldsymbol R}^{} \rangle_{\text A}^{}
                    + | \hexagon_{\boldsymbol R}^{} \rangle_{\text B}^{} 
                   \big],
\end{eqnarray}
where ${\boldsymbol R}$ refers to the position of the resonating hexagons,
and A and B label the two charge occupation configurations of the three 
electons on the resonating hexagon (see Fig.~\ref{fig1}). The spin quantum 
number can still be transferred via the spin exchange interaction, so 
${| \Psi \rangle_{\text c}^{} }$ merely represents the charge wavefunction.

\section{The emergent orbitals and the local moments}
\label{sec3}

In this section, we focus on the strong plaquette charge ordered 
regime and reveal the novel features of the local moment structure. 
The three electrons are well localized on the resonating hexagons 
but still move in the collective fashion that is governed by $H_{\text{eff}}$. 
This collective motion tunnels the electron spins that are interacting 
with the superexchange interaction at the same time. As a comparison, 
the localized electrons on a lattice site of a conventional Mott insulator 
are fully governed by the atomic electron interactions and the Hund's rules. 
Here, the right model that describes the localized electrons on an 
individual resonating hexagon is
\begin{eqnarray}
H_{\hexagon_{\boldsymbol R}}^{} 
&=& - 
{K}_1 \sum_{\alpha \beta \gamma } (c^{\dagger}_{1\alpha} c^{\phantom\dagger}_{6\alpha} 
                                 c^{\dagger}_{5\beta} c^{\phantom\dagger}_{4\beta}  
  c^{\dagger}_{3\gamma} c^{\phantom\dagger}_{2\gamma}    + h.c.) 
\nonumber \\
&  & - 
{K}_2  \sum_{\alpha \beta \gamma } (c^{\dagger}_{1\alpha} c^{\phantom\dagger}_{2\alpha} 
                                 c^{\dagger}_{3\beta} c^{\phantom\dagger}_{4\beta}  
                                  c^{\dagger}_{5\gamma} c^{\phantom\dagger}_{6\gamma} 
+ h.c.) 
\nonumber \\
&& + H_{\text{ex},{\boldsymbol R}},
\end{eqnarray}
where the superexchange interaction is given as
\begin{eqnarray}
H_{\text{ex},{\boldsymbol R}} = J \sum_{\langle\langle ij\rangle\rangle 
\in \hexagon^{}_{\boldsymbol R}}  
 ({\bf S}_i \cdot {\bf S}_j - \frac{1}{4} ) \, n_i n_j  . 
 \label{eq4}
\end{eqnarray} 
It is interesting to note that the above superexchange differs 
from the usual form of the exchange interaction by having extra 
electron density opertors $n_i$ and $n_j$. This is because the
positions of the electrons are not fixed due to their collective tunneling
on the hexagon plaquette. The local Hilbert space of 
$H_{\hexagon_{\boldsymbol R}}^{}$ also differs significantly from
the on-site one for a conventional Mott insulator, and is instead 
spanned by the electron states that are labelled by both
the positions and the spin quantum numbers of the three resonating electrons. 
Because the electrons are separated from each other by one lattice site 
due to the repulsive interaction, the Hilbert space for the electron 
positions is highly constrained. For the resonating hexagon
centered at $\boldsymbol{R}$, there are in total 16 states that are labelled by
\begin{eqnarray}
 | \alpha\beta\gamma\rangle_{\text{A}} &\equiv& |n_1=0\rangle |n_2=1,\alpha \rangle
                  |n_3 = 0 \rangle 
                  \nonumber \\
&\times &  |n_4=1,\beta \rangle | n_5 =0\rangle |n_6=1, \gamma \rangle , 
\\
|\alpha\beta\gamma \rangle_{\text{B}} &\equiv& | n_1=1, \alpha\rangle 
                    |n_2 =0\rangle |n_3=1,\beta\rangle
\nonumber \\
&\times&  |n_4=0 \rangle |n_5=1,\gamma\rangle |n_6 =0 \rangle  , 
\end{eqnarray}
where $\alpha,\beta,\gamma$ ($= \uparrow, \downarrow$) 
refer to the electron spins at the occupied site.  
Since the hexagonal Hamiltonian $H_{\hexagon_{\boldsymbol R}}^{}$ 
commutes with the total spin ${\bf S}_{\text{tot}}$ and $S^z_{\text{tot}}$
of the three resonating electrons, we use $\{ {S}_{\text{tot}},  
S^z_{\text{tot}}\}$ to label the spin states of the hexagon plaquette. 
From the spin composition rule for three electron spins, we have the
following relation 
\begin{eqnarray}
\frac{1}{2}\otimes \frac{1}{2} \otimes\frac{1}{2} 
\equiv
\frac{1}{2} \oplus \frac{1}{2} \oplus \frac{3}{2} ,   
\end{eqnarray}
where the left side are the product state of the three electron spins
and the right side are the total spin states ${S}_{\text{tot}}$. 
For both A and B occupation configurations, there are eight 
spin states. Note that we have two pairs of ${{S}_{\text{tot}} =1/2}$ 
states for each occupation configuration.

The two states with ${S_{\text{tot}} = {3}/{2}}$ are 
simply the ferromagnetic states and are certainly not
favored by the antiferromagnetic exchange interaction 
$H_{\text{ex},\boldsymbol{R}}$. Directly solving the 
Hamiltonian $H_{\hexagon_{\boldsymbol R}}$, 
we find that when 
\begin{equation}
J > \frac{2}{3} \big[  K_1 +   K_2 
- (  K_1^2 -  K_1   K_2 +  K_2^2  )^{\frac{1}{2}} \big] ,
\end{equation} 
the local ground states are four symmetric states with ${S_{\text{tot}} =1/2}$.
Here, the ``symmetric'' states refer to being symmetric between the A and B occupation 
configurations in Fig.~\ref{fig3}. This is understood by the fact that 
the collective motion of three electrons favors symmetric states 
rather than antisymmetric ones. 
These four-fold degenerate states can be effectively characterized 
by two quantum numbers $\{s^z,\tau^z\}$ with  ${s^z = \pm \frac{1}{2}}$ 
and ${\tau^z = \pm \frac{1}{2}}$, where $s^z$ refers to the total spin 
${s^z \equiv S^z_{\text{tot}} = \pm \frac{1}{2}}$. The pseudospin-${1}/{2}$ 
operator $\boldsymbol{\tau}$ refers to the emergent orbitals that 
will be explained below.

\begin{figure}[t]
{\includegraphics[width=8cm]{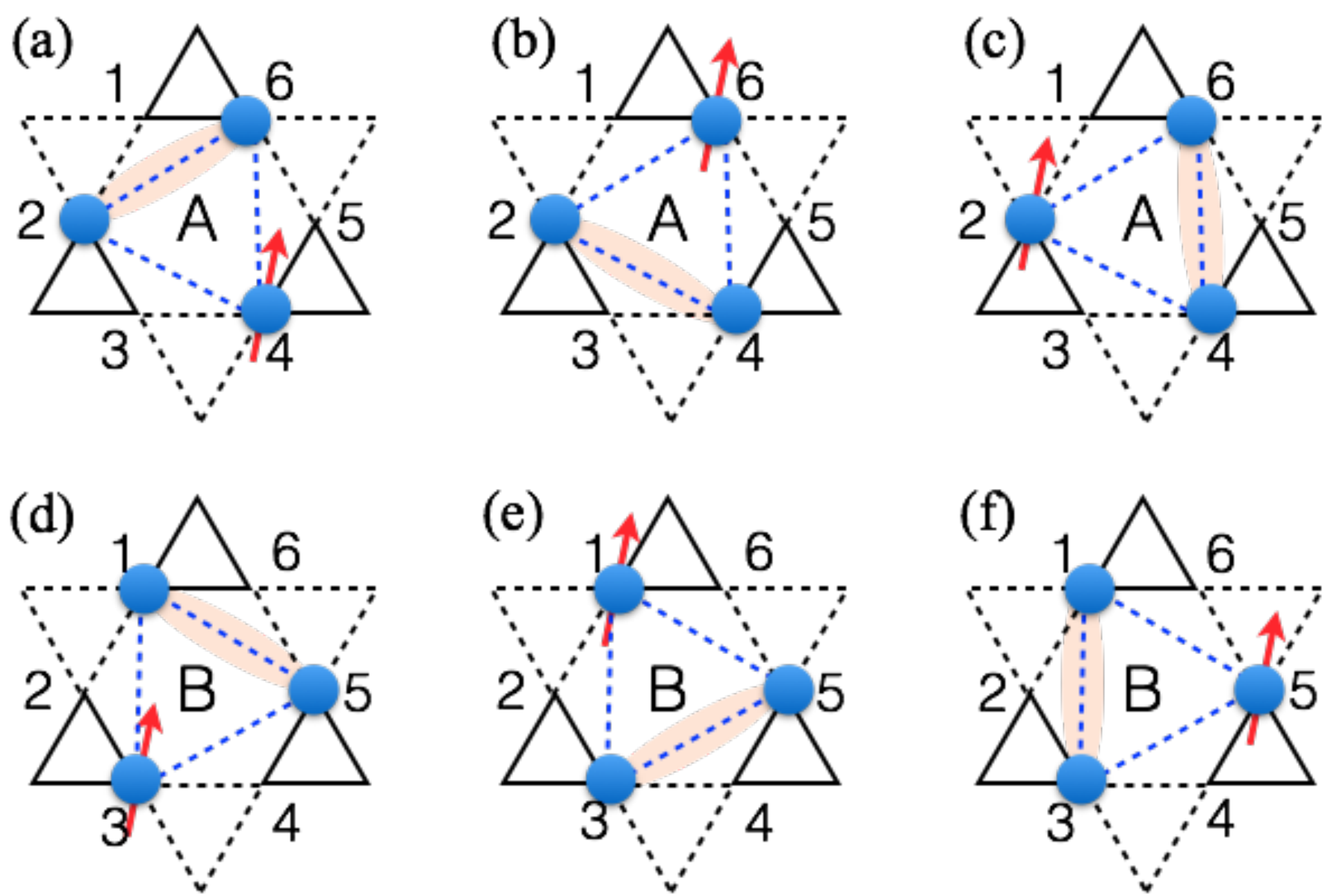}}
\caption{(Color online.) 
Three spin-singlet positions for both A and B occupation configurations.
The (orange) dimer 
refers to the spin singlet, and the (red) arrow is the dangling spin.
The three spin-singlet configurations are related by the three-fold rotation
around the hexagon center.  }
\label{fig3}
\end{figure}

The wavefunctions of the four degenerate states 
are labelled by 
$| {\tau^z s^z} \rangle^{}_{\boldsymbol R}$ 
and are given as 
(to the order of $\mathcal{O} ({{K}_2}/{{K}_1})$~\footnote{We 
expect that ${V_1\gtrsim V_2, t_1>t_2}$, so ${K_1 \gg K_2}$ for 
LiZn$_2$Mo$_3$O$_8$.}),
\begin{eqnarray}
|{\uparrow \uparrow }\rangle_{\boldsymbol R}^{} &=& \frac{1}{2} \big[
 |{\uparrow\uparrow\downarrow} \rangle_{\text A}^{} 
 - |{\uparrow\downarrow\uparrow} \rangle_{\text A}^{} \nonumber \\
 && \,\,
+ | {\downarrow \uparrow\uparrow}\rangle_{\text B}^{}
- |{\uparrow\uparrow\downarrow} \rangle_{\text B}^{} 
  \big],
\label{eqa}
\\
 |{\downarrow\uparrow} \rangle_{\boldsymbol R}^{} &=& \frac{\sqrt{3}}{6} \big[
 2 |{ \downarrow \uparrow\uparrow}
 \rangle_{\text A}^{}
 - | {\uparrow\downarrow \uparrow} \rangle_{\text A}^{}
 - | {\uparrow \uparrow\downarrow} \rangle_{\text A}^{}
  \nonumber \\
  && \quad\,  + 2|  {\uparrow\downarrow\uparrow} \rangle_{\text B}^{} 
   - | { \uparrow \uparrow\downarrow} \rangle_{\text B}^{}
   - | { \downarrow \uparrow\uparrow} \rangle_{\text B}^{} \big],
\label{eqb}
\end{eqnarray}
and these other two states $|{\uparrow \downarrow} \rangle_{\boldsymbol R}^{}$, 
$|{\downarrow \downarrow} \rangle_{\boldsymbol R}^{}$ 
are simply obtained by applying a time-reversal transformation to the above 
two states,
\begin{eqnarray}
|{\uparrow \downarrow } \rangle_{\boldsymbol R}^{} 
&=& \frac{1}{2} \big[
 |{\downarrow\downarrow\uparrow} \rangle_{\text A}^{} 
 - |{\downarrow\uparrow\downarrow} \rangle_{\text A}^{} \nonumber \\
 && \,\,
+ | {\uparrow \downarrow\downarrow}\rangle_{\text B}^{}
- |{\downarrow\downarrow\uparrow} \rangle_{\text B}^{} 
  \big],
\label{eqc}
\\
 |{\downarrow\downarrow} \rangle_{\boldsymbol R}^{} &=& \frac{\sqrt{3}}{6} \big[
 2 |{ \uparrow \downarrow\downarrow}
 \rangle_{\text A}^{}
 - | {\downarrow\uparrow \downarrow} \rangle_{\text A}^{}
 - | {\downarrow \downarrow\uparrow} \rangle_{\text A}^{}
  \nonumber \\
  && \quad\,  + 2|  {\downarrow\uparrow\downarrow} \rangle_{\text B}^{} 
   - | { \downarrow \downarrow \uparrow} \rangle_{\text B}^{}
   - | { \uparrow \downarrow\downarrow} \rangle_{\text B}^{} \big].
\label{eqd}
\end{eqnarray}

We clarify the physical origin of the four-fold degeneracy of the above
four states for the hexagon plaquette. First, the two-fold degeneracy of 
${s^z =\pm 1/2}$ is simply protected by the time-reversal symmetry. 
The remaining two-fold degeneracy comes from the point group symmetry of 
the resonating hexagon. This is ready to see if we fix the occupation 
configuration of the three electrons. To be more specific, let us start 
with the A configuration in the upper panel of Fig.~\ref{fig3}. 
To optimize the antiferromagnetic exchange interaction, 
two electron spins out of the three must form a spin singlet, 
leaving the third electron as a dangling spin. 
As shown in Fig.~\ref{fig3}, the spin singlet can be formed 
between any pair of the electrons, and the different arrangements 
of the spin singlet are related by the three-fold rotation. 
Although there seems to be three possible singlet arrangements, 
only two of them are linearly independent and are responsible for 
the two-fold degeneracy. Likewise, for the B configuration on the 
lower panel of Fig.~\ref{fig3}, we again have two such degenerate 
states. When the three electrons start to move collectively within 
the hexagon between the A and B configurations, 
the corresponding states start to hybridize and 
the symmetric states are favored energetically. 
The two-fold degeneracy survives and is given as 
the ${\tau^z = \uparrow,\downarrow} $ states in 
Eqs.~\eqref{eqa}-\eqref{eqd}.  

The three electrons are localized on the resonating hexagon 
but are delocalized within the resonating hexagon. It is hard 
for them to move out of the resonating hexagon, but easy for 
them to move within the resonating hexagon. Due to this collective 
motion, the wavefunctions of $|\tau^z s^z\rangle$ are extended 
and span across the resonating hexagon, and the $\tau^z=\uparrow,\downarrow$ 
states behave like two degenerate orbitals that are defined on 
the resonating hexagon. Since the degeneracy of   
${\tau^z = \uparrow,\downarrow}$ states originate 
from the arrangements of the spin singlets, the pseudospin 
$\boldsymbol{\tau}$ is {\it even} under the time-reversal 
transformation. The two emergent orbital states that are defined 
in Eqs.~\eqref{eqa} and \eqref{eqb} comprise the two-dimensional 
E-type irreducible representation of the point group, and thus their 
two-fold degeneracy is protected by the point group symmetry of 
the resonating hexagon.

\section{The Kugel-Khomskii spin-orbital interaction}
\label{sec4}

In this section we study and derive the interaction between the spins 
and the emergent orbitals that live on the neighboring resonating 
hexagons. This interaction is necessarily of the Kugel-Khomskii type.  
Based on the Kugel-Khomskii model, we obtain the Curie-Weiss 
temperature and Curie constant in the strong plaquette ordered 
regime, and compare with the high temperature results. 

\subsection{The Kugel-Khomskii model}
\label{sec4a}

The neighboring resonating hexagons are connected by a ``bow-tie'' 
structure that is composed of the corner-shared up and down triangles 
(see Fig.~\ref{fig4}). The local moment interaction comes from 
the remaining exchange interaction between the two electron spins 
that reside on the four exterior vertices of the bow-tie. 
To illustrate the idea, we consider the bow-tie structure that 
connects the two resonating hexagons centered at ${\boldsymbol R}$ 
and ${{\boldsymbol R} + {\boldsymbol a}_1}$ (see Fig.~\ref{fig1} 
and Fig.~\ref{fig4}). To derive the local moment interaction, 
we need to project the remaining electron spin exchange interaction 
onto the four-fold degenerate local moment states $|\tau^zs^z\rangle$ 
of each resonating hexagon. For this purpose, we first write down 
the inter-hexagon exchange interaction between the electrons at 
the bow-tie vertices,
\begin{eqnarray}
H'_{\text{ex}} &=& 
 - \frac{J\rq{}}{4} 
[  n_4 ( {\boldsymbol R} )  +   n_5 ({\boldsymbol R})  ]
[  n_1 (  {\boldsymbol R} + {\boldsymbol a}_1 )  
 + n_2 ({\boldsymbol R} + {\boldsymbol a}_1)]
\nonumber \\
&& +
J\rq{} [ {\boldsymbol S}_{4}( {\boldsymbol R}) n_{4} 
({\boldsymbol R})  + {\boldsymbol S}_{5}({\boldsymbol R}) n_{5} ({\boldsymbol R})  ]
 \cdot [ {\boldsymbol S}_{1} ({\boldsymbol R}+{\boldsymbol a}_1) 
\nonumber \\
&& \quad \,\, \times 
n_{1} ({\boldsymbol R}+{\boldsymbol a}_1)  + {\boldsymbol S}_{2} 
({\boldsymbol R}+{\boldsymbol a}_1) n_{2} ( {\boldsymbol R}+{\boldsymbol a}_1 ) ],
\label{eq7}
\end{eqnarray}
where we have included the exchange interactions for electrons at 
all four pairs of the external vertices. Again, since the position of the electron is not fixed,
the electron number operator $n_i$ is introduced. The exchange paths
all go through the central vertex of the bow-tie and are of equal length. Therefore,
only one exchange coupling $J'$ is introduced for all the four pairs in Eq.~\eqref{eq7}. 
The exchange coupling $J'$ can be obtained from the fourth order perturbation theory and 
is given as
\begin{eqnarray}
J' & = & 
  \frac{4 t_1^2 t_2^2 }{ U V_1^2 } 
+ \frac{4 t_1^2 t_2^2 }{ U V_2^2 }
+ \frac{4 t_1^2 t_2^2 }{ U V_1 V_2 }, 
\end{eqnarray}
and the fifth order perturbation theory could introduce more terms to $J'$ 
without invoking double electron occupancy on a single lattice site. 
Moreover, since $J\rq{}$ is the exchange coupling between the spins 
in the strong plaquette ordered regime, $J'$ is expected to be  
weaker than the intra-resonating-hexagon exchange coupling $J$ 
in Eq.~\eqref{eq4}.

\begin{figure}[t]
{\includegraphics[width=6.4cm]{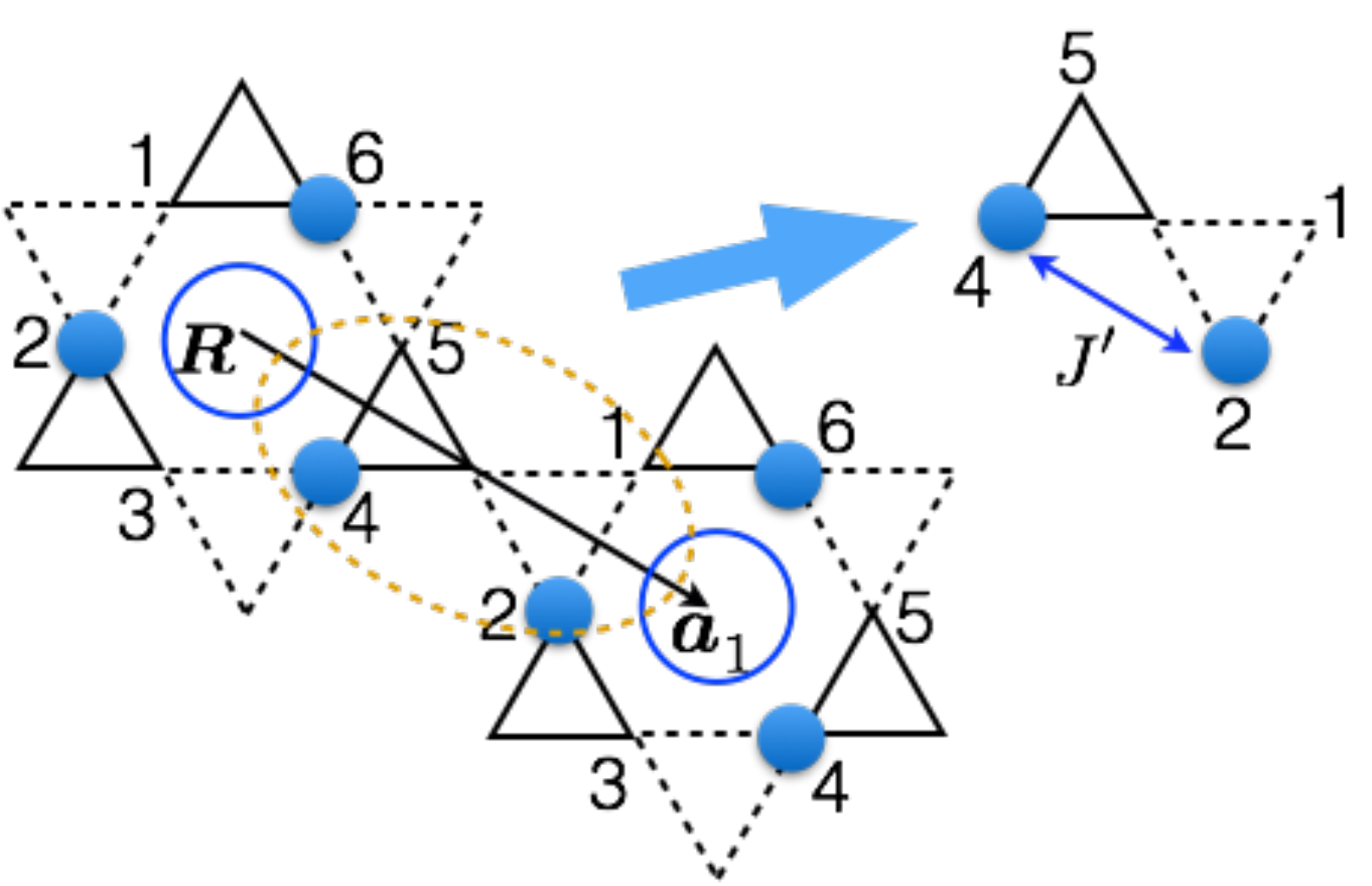}}
\caption{(Color online.) The bow-tie structure that connects 
two neighboring resonating hexagons. In the upper right corner, 
we indicate the exchange interaction $J'$ between two electrons. 
}
\label{fig4}
\end{figure}

We project $H'_{\text{ex}}$ onto the local ground state manifold
at resonating hexagon sites ${\boldsymbol R}$ and 
${\boldsymbol R}+{\boldsymbol a}_1$
and then express the resulting interaction in terms of the spin 
$\boldsymbol{s}$ and the pseudospin $\boldsymbol{\tau}$. 
The effective interaction on other bonds can be obtained likewise. 
The final local moment interaction reduces to a 
Kugel-Khomskii model~\cite{Kugel82} that is defined on the 
triangular lattice formed by the resonating hexagons.  
To the order of $\mathcal{O}( { K_2}/{ K_1} )$, the Kugel-Khomskii model
is given as
\begin{eqnarray}
H_{\text{KK}} & = & \frac{J'}{9} 
 \sum_{{\boldsymbol R}} 
 \sum_{\mu=x,y,z}  
( {\boldsymbol s}_{\boldsymbol R}^{} \cdot 
 {\boldsymbol s}_{{{\boldsymbol R}+{\boldsymbol a}_{\mu} } }^{}   )
 \nonumber \\
 &&  \quad \times 
[  1 +4  T^{\mu}_{\boldsymbol R}  ]
[  1 - 2 T^{\mu}_{{\boldsymbol R}+{\boldsymbol a}_{\mu}}  ], 
\label{eq9}
\end{eqnarray}
where the new set of pseudospin operators, $T^{\mu}$'s, are defined as 
\begin{eqnarray}
T^{x}_{\boldsymbol R} & = & -\frac{1}{2} \tau^z_{\boldsymbol R} -  
\frac{\sqrt{3}}{2} \tau^x_{\boldsymbol R} ,
\\
T^{y}_{\boldsymbol R} & = & -\frac{1}{2} \tau^z_{\boldsymbol R} +  
\frac{\sqrt{3}}{2} \tau^x_{\boldsymbol R} ,\\
T^z_{\boldsymbol R}   & = & \tau^z_{\boldsymbol R},
\end{eqnarray}
and ${{\boldsymbol a}_{x} = {\boldsymbol a}_{1}}, {{\boldsymbol a}_y = {\boldsymbol a}_2}$ 
and $ {{\boldsymbol a}_z =- {\boldsymbol a}_1 - {\boldsymbol a}_2 }$. 
The particular expression of the Kugel-Khomskii model in Eq.~\eqref{eq9} originates
from the choice of two orbital wave functions in Eqs.~\eqref{eqa}-\eqref{eqd}. 
If a different set of orbital wavefunctions is chosen, the 
resulting Kugel-Khomskii model would have a different form. 
In Eq.~\eqref{eq9}, the effective exchange coupling is significantly 
reduced by after the projection compared to the original exchange 
coupling $J'$ in Eq.~\eqref{eq7}. 
The important factor 1/9 in front of this equation can be understood 
physically as coming from the fact that each spin is found in the 
bowtie structure connecting two hexagons only 1/3 of the time.

\subsection{The Curie-Weiss laws}
\label{sec4b}

Since the pseudospin $\boldsymbol{\tau}$ is even under the time reversal 
transformation and thus does not couple to the external magnetic field, 
the low-temperature Curie-Weiss temperature thus detects the spin-spin 
interaction, and from the Kugel-Khomskii model $H_{\text{KK}}$ 
we directly compute the Curie constant $\mathcal{C}$ and the Curie-Weiss 
temperature $\Theta_{\text{CW}}$ at the low temperature,
\begin{eqnarray}
&& \mathcal{C}^{\text L}  = \frac{g^2 \mu_{\text B}^2 s(s+1) }{3k_{\text B}} 
                            \frac{N}{3}, \\
&& \Theta_{\text{CW}}^{\text L}  = - \frac{2 s(s+1)J'}{9} ,  
\label{cwl} 
\end{eqnarray}
where $N$ is the total number of electrons, $g$ is the Land\'{e} factor, 
and ``${N}/{3}$'' in $\mathcal{C}^{\text L}$ means the active spin degrees of 
freedom in the strong plaquette ordered phase comprise 1/3 of the 
total number of electrons. This is a natural consequence due to the 
spin state reconstruction within each resonating hexagons. This result
is consistent with the low-temperature magnetic susceptibility
LiZn$_2$Mo$_3$O$_8$~\cite{Mourigal14,Sheckelton14,Sheckelton12,Mcqueen2015}. 

To make a comparison with the high-temperature susceptibility,
we consider the high temperature regime where the plaquette 
charge order is present and the spin singlet within the resonating
hexagon plaquette is thermally destroyed. In this regime, 
all the electron spins contribute to the magnetic susceptibility. 
Therefore, the Curie constant for this high-temperature regime
is simply given by
\begin{equation}
\mathcal{C}^{\text H}  = 
\frac{g^2 \mu_{\text B}^2 s(s+1) }{3k_{\text B}} N,
\end{equation}
and is three times larger than the low-temperature one. 
Moreover, in this regime, Fig.~\ref{fig1} is a typical electron configuration. 
For each electron, there are four neighboring electrons
that interact with this electron spin with the
pairwise spin interaction across the bow-tie structure.  
Among these interactions, there are two intra-resonating-hexagon interactions
with the coupling $J$ and two inter-resonating-hexagon
interaction $J'$. Then the Curie-Weiss temperature
for this high-temperature regime is given as
\begin{eqnarray}
\Theta_{\text{CW}}^{\text H}  = - \frac{2 s(s+1)}{3} (J+J'),
\label{cwl} 
\end{eqnarray}
and is $3(1+J/J')$ times larger than the low-temperature one. 
Since $J'$ is expected to be less than $J$, 
the ratio is larger than 6 and provides a separation of 
scale between the high temperature freezing of 2/3 of the 
spins and the interaction among the remaining spins.
In the experiment on LiZn$_2$Mo$_3$O$_8$, the two Curie-Weiss 
temperatures are $-220$K and $-14$K respectively~\cite{Sheckelton12}.

\section{Parton construction for the candidate spin liquid state}
\label{sec5}

Like any other spin-orbital exchange models~\cite{Kugel82}, 
the Kugel-Khomskii model $H_{\text{KK}}$ in our context involves 
the spin-spin interaction, the pseudospin-pseudospin 
interaction and the spin-pseudospin interaction,
and all these interactions are of the same energy scale.  
These interactions together make the model analytically intractable. 
In the absence of the spin-pseudospin interaction, 
the Heisenberg spin exchange model would favor the conventional 
120-degree state with a long-range order. The spin-pseudospin interaction, 
however, competes with the Heisenberg term, destabilizes the
conventional 120-degree state and may potentially induce a spin liquid 
state. This is because the quartic-like spin-pseudospin interaction
allows the local moment to fluctuate more effectively in the spin-pseudospin 
space. Such a spin liquid, if exists, may be smoothly connected to 
the U(1) spin liquid with spinon Fermi surfaces that was proposed for the 
weak plaquette charge ordered regime in Ref.~\onlinecite{ChenPRB2016}. 

From the experimental side, a broad continuous excitation has been
discovered in the inelastic neutron scattering measurement on
powder samples. The authors in Ref.~\onlinecite{Mourigal14} 
proposed a gapless spin liquid state. Moreover, the neutron spectral 
weight in the experiment is not suppressed at low energies, which 
indicates that the ground state cannot be a Dirac spin liquid. 
Based on the experimental results, we here propose 
the candidate ground state to a spin liquid with a spinon 
Fermi surface. This phenomenological proposal is again consistent with 
the previous suggestion from the weak coupling approach~\cite{ChenPRB2016}. 
To demonstrate the phenomenological consequence of this proposal, 
we develop a new parton construction that is designed for our 
spin-orbital model and suggest the experimental consequence 
of this candidate state.

\subsection{The parton construction} 
\label{secsuba}

There are both spin and orbital degrees of freedom on a single site ${\boldsymbol R}$.  
To account for both of them, we introduce the following fermionic parton representation,
\begin{eqnarray}
\boldsymbol{\tau}_{\boldsymbol R}^{} & = & 
\sum_{m,n} \sum_{\alpha}
\frac{1}{2} f^{\dagger}_{{\boldsymbol R}m\alpha} 
\boldsymbol{\sigma}_{mn}^{} f^{}_{{\boldsymbol R}n\alpha},
\\
\boldsymbol{s}_{\boldsymbol R}^{} & = &
\sum_{m} \sum_{\alpha,\beta} 
\frac{1}{2} f^{\dagger}_{{\boldsymbol R}m\alpha} 
\boldsymbol{\sigma}_{\alpha\beta}^{} f^{}_{{\boldsymbol R}m\beta},
\end{eqnarray}
where $m,n = \,\uparrow,\downarrow$ refer to the pseudospin state for the 
orbitals, $\alpha,\beta =\, \uparrow,\downarrow$ refer to the spin state, 
and $\boldsymbol{\sigma} = (\sigma^x,\sigma^y,\sigma^z)$ is the vector of 
Pauli matrices. To get back to the physical Hilbert space, we further impose 
a Hilbert space constraint
\begin{eqnarray} 
\sum_{\alpha} \sum_{m}
f^{\dagger}_{{\boldsymbol R}m\alpha}  f^{}_{{\boldsymbol R}m\alpha} = 1.
\end{eqnarray}
Unlike the pure spin model, our spinon carries an extra orbital index. 
This parton construction could be well extended to other spin-orbital 
models. 

\begin{figure}[t]
{\includegraphics[width=8cm]{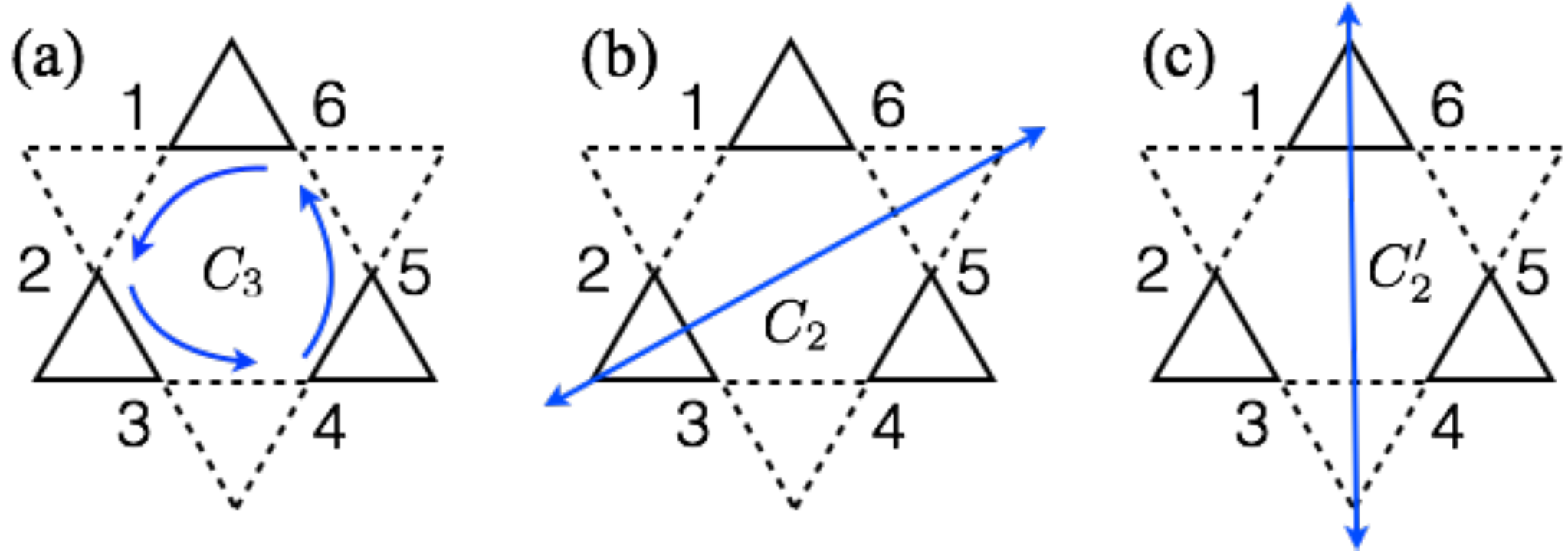}}
\caption{(a) Three fold rotation around the center of the plaquette.
(b) The two fold rotation axis.
(c) Another two fold rotation axis. 
}
\label{figsym}
\end{figure}

\subsection{The symmetry property of the spinons}
\label{secsubb}

The U(1) spin liquid with 
spinon Fermi surfaces was proposed for the weak plaquette 
charge ordered regime~\cite{ChenPRB2016}. For this state, 
the spinon transforms identically like the electron under 
the lattice transformation, 
and there is no projective realization of the lattice symmetry. 
Since we suggest that the possible spin liquid for our Kugel-Khomskii
model in the strong coupling regime is connected to the ground state
in the weak plaquette ordered regime, we here explictly derive the
symmetry transformation of the spinons in our context. 

Let us consider a single plaquette at ${\boldsymbol R}$, the symmetries
include the three-fold rotation $C_3$ and two two-fold rotations $C_2$ 
and $C_2'$ (see Fig.~\ref{figsym}). The lattice symmetry does not 
change the spin component, but acts on the orbital degree of freedom.  
Under $C_3$, the lattice sites within the hexagon plaquette transform 
as
\begin{eqnarray}
C_3: && \quad 2\rightarrow 4,\, 4\rightarrow 6,\, 6\rightarrow 2 , \\
C_3: && \quad 1\rightarrow 3,\, 3\rightarrow 5,\, 5\rightarrow 1 , 
\end{eqnarray}
therefore, from the orbital wavefunctions, we have that 
the states $|{\uparrow\uparrow} \rangle_{\boldsymbol R}$
and $|{\downarrow\uparrow}\rangle_{\boldsymbol R}$ transform as
\begin{eqnarray}
C_3: \quad |{\uparrow\uparrow}\rangle_{\boldsymbol R}   &\rightarrow & 
 -\frac{1}{2} |{\uparrow\uparrow}\rangle_{\boldsymbol R} 
 + \frac{\sqrt{3}}{2} |{\downarrow\uparrow}\rangle_{\boldsymbol R},
\\
C_3: \quad |{\downarrow\uparrow}\rangle_{\boldsymbol R} &\rightarrow & 
-\frac{\sqrt{3}}{2} |{\uparrow\uparrow}\rangle_{\boldsymbol R}
-\frac{1}{2} |{\downarrow\uparrow}\rangle_{\boldsymbol R},
\end{eqnarray}
where the transformation does not depend on the spin quantum number,
and the identical transformations are obtained for the states 
$|{\uparrow\downarrow} \rangle_{\boldsymbol R}$ and 
$|{\downarrow\downarrow}\rangle_{\boldsymbol R}$.
One then establishes 
\begin{eqnarray}
C_3: \quad f_{{\boldsymbol R}\uparrow\alpha}   & \rightarrow & 
- \frac{1}{2} f_{{\boldsymbol R}\uparrow\alpha}
+ \frac{\sqrt{3}}{2} f_{{\boldsymbol R}\downarrow\alpha} ,
\\
C_3: \quad f_{{\boldsymbol R}\downarrow\alpha} & \rightarrow & 
-\frac{\sqrt{3}}{2} f_{{\boldsymbol R}\uparrow\alpha}
- \frac{1}{2} f_{{\boldsymbol R}\downarrow\alpha}. 
\end{eqnarray}
Following the same type of calculation, under $C_2$ and $C_2'$, 
we have 
\begin{eqnarray}
C_2:  \quad 5\leftrightarrow 6,\, 1 \leftrightarrow 4, \, 2 \leftrightarrow 3,
\end{eqnarray}
and 
\begin{eqnarray}
C_2^{}:\quad && f_{{\boldsymbol R}\uparrow\alpha}^{} 
\rightarrow - f_{{\boldsymbol R}\uparrow\alpha}^{}  , 
\\
C_2^{}:\quad && f_{{\boldsymbol R}\downarrow\alpha}^{} 
\rightarrow + f_{{\boldsymbol R}\downarrow\alpha}^{},
\end{eqnarray}
and 
\begin{eqnarray}
C_2': \quad 1 \leftrightarrow 6, 2\leftrightarrow 5, 3 \leftrightarrow 4, 
\end{eqnarray}
and 
\begin{eqnarray}
C_2':\quad && f_{{\boldsymbol R}\uparrow\alpha}^{} 
\rightarrow + \frac{1}{2} f_{{\boldsymbol R}\uparrow\alpha}^{} 
            -\frac{\sqrt{3}}{2}f_{{\boldsymbol R}\downarrow\alpha}^{}   , 
\\
C_2':\quad && f_{{\boldsymbol R}\downarrow\alpha}^{} 
\rightarrow - \frac{\sqrt{3}}{2} f_{{\boldsymbol R}\uparrow\alpha}^{} 
            -\frac{1}{2}f_{{\boldsymbol R}\downarrow\alpha}^{} .
\end{eqnarray}

\subsection{The spinon Fermi surface state} 
\label{secsubc}

From the spinon symmetry properties, we determine the generic 
symmetry allowed spinon mean-field Hamiltonian $H_{\text{spinon}}$,
\begin{eqnarray}
H_{\text{spinon}} & = & 
\sum_{{\boldsymbol R},\mu}
\sum_{m,n} \sum_{\alpha}  
{t^{\mu}_{mn}} 
f^{\dagger}_{{\boldsymbol R}m\alpha} 
f^{}_{{\boldsymbol R}+{\boldsymbol a}_{\mu},n,\alpha}
+ h.c.  ,
\end{eqnarray}
where $t^{\mu}_{mn}$ is a bond dependent hopping matrix
for the spinons, and we have the symmetry allowed hoppings as
\begin{eqnarray}
t^x & = & -
{\tilde t}_1 \mathbb{1}_{2\times 2} + {\tilde t}_2 \sigma^z + \sqrt{3} {\tilde t}_2 \sigma^x ,
\\
t^y & = & -
{\tilde t}_1 \mathbb{1}_{2\times 2} + {\tilde t}_2 \sigma^z - \sqrt{3} {\tilde t}_2 \sigma^x ,
 \\
t^z & = & - 
{\tilde t}_1 \mathbb{1}_{2\times 2} - 2{\tilde t}_2 \sigma^z ,
\end{eqnarray}
and $\mathbb{1}_{2\times 2}$ is a $2\times 2$ identity matrix. 
This model describes the spinon hopping on the triangular lattice 
with two orbitals at each lattice site. Since the spinons are at 
1/4 filling, each band is partially filled and the system develops spinon Fermi 
surfaces (see Fig.~\ref{figspinon}). 
The mean-field ground state is obtained by filling the 
spinon states below the Fermi energy $E_{\text F}$,
\begin{eqnarray}
 | \Psi_{\text{MF}} \rangle  
 = \prod_{E_{{\boldsymbol k},j} < E_{\text F}} \xi_{ {\boldsymbol k}j\uparrow}^\dagger
 \xi_{ {\boldsymbol k}j\downarrow}^\dagger |0\rangle,
\end{eqnarray}
where $E_{{\boldsymbol k},j}$ is the energy of the eigenmode 
that is defined by $\xi_{ {\boldsymbol k}j\uparrow}^\dagger$
or $\xi_{ {\boldsymbol k}j\downarrow}^\dagger$,
and is given as
\begin{eqnarray}
E_{{\boldsymbol k},1} &=& - 2 {\tilde t}_1 (c_x+c_y+c_z) + 4 |{\tilde t}_2| (c_x^2+c_y^2+c_z^2
\nonumber \\
&&  \quad\quad\quad\quad\quad\quad\quad -c_yc_z-c_xc_z-c_xc_y )^{\frac{1}{2}} ,
\\
E_{{\boldsymbol k},2} &=& -2 {\tilde t}_1 (c_x+c_y+c_z) - 4 |{\tilde t}_2| (c_x^2+c_y^2+c_z^2 
\nonumber \\
&&  \quad\quad\quad\quad\quad\quad\quad -c_yc_z-c_xc_z-c_xc_y )^{\frac{1}{2}} .
\end{eqnarray}
Here ${c_{\mu} = \cos ( {\boldsymbol k}\cdot {\boldsymbol a}_{\mu} ) }$.

\begin{figure}[t]
{\includegraphics[width=6.5cm]{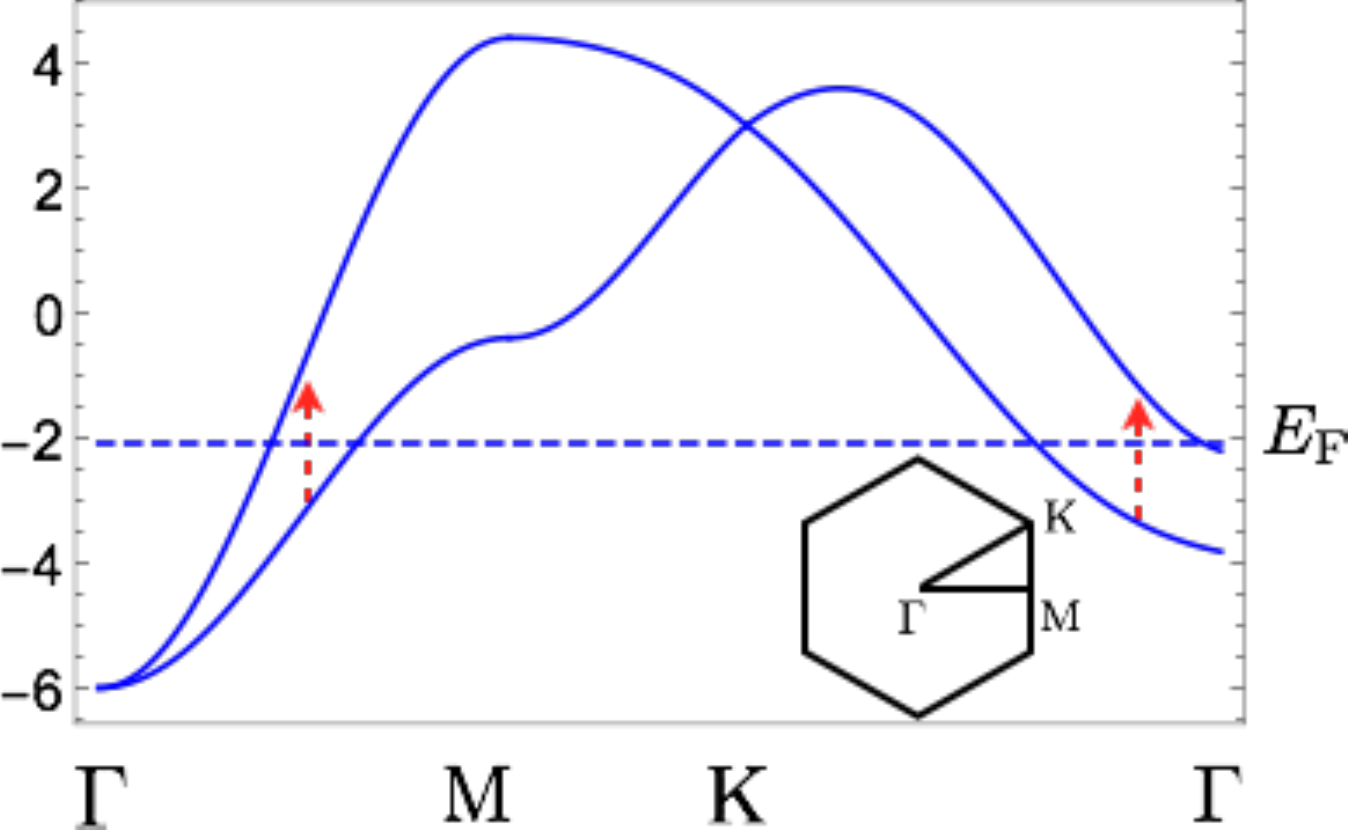}}
\caption{(Color online.) The two spinon bands and their vertical particle-hole 
transition between the two bands. In the plot, ${{\tilde t}_2 = 0.3 {\tilde t}_1}$, 
and ${ {\tilde t}_1=1}$ is used as the energy unit. The inset is the Brillouin 
zone of the triangular lattice formed by the reonating hexagons.}
\label{figspinon}
\end{figure}

\subsection{The qualitative feature of the spinon continuum due to the emergent orbitals} 
\label{secsubd}

The key property for the spinon mean-field model is the presence of the 
inter-orbital hopping ${\tilde t}_2$ that hybridizes different orbitals 
such that each spinon band no longer has a definite orbital character. 
This inter-orbital spinon hopping arises from the fact that the orbital 
interaction is anisotropic in the orbital space and only respects 
the discrete lattice symmetry. In the inelastic neutron scattering, 
the neutron would only see the effective spin and not see the emergent orbital
degree of freedom. The orbital degree of freedom, however, 
has an important effect on the spinon continuum that is observed by 
inelastic neutron scattering. The neutron detects the particle-hole 
excitation across the spinon Fermi level. From the momentum and energy 
conservation, we have the momentum and energy transfer of the neutron as 
\begin{eqnarray}
{\boldsymbol q} &=& {\boldsymbol q}_1 - {\boldsymbol q}_2 ,\\
E &=& E_{{{\boldsymbol q}_1},j_1} - E_{{{\boldsymbol q}_2},j_2} ,
\end{eqnarray}
where ${\boldsymbol q}_1$ and $E_{{{\boldsymbol q}_1},j_1}$
are the momentum and the energy of an unoccupied spinon
while ${\boldsymbol q}_2$ and $E_{{{\boldsymbol q}_2},j_2}$
are the momentum and the energy of the filled spinon.
The particle-hole excitation would involve both 
the intra-band transitions (with $j_1=j_2$) and the inter-band 
transitions (with $j_1\neq j_2$). If there is no orbital degree of 
freeom and there is only one single spinon band, the inter-band 
transition is not involved, and the inelastic neutron scattering 
spectral weight for the intra-band transition is suppressed for 
the finite energies at the $\Gamma$ point. This is because at the 
mean-field level the intra-band process always excites the 
finite-energy spinon particle-hole pair with a finite 
momentum. In contrast, with the inter-band vertical process 
(see Fig.~\ref{figspinon}), the spinon particle-hole pair with 
zero momentum can carry a wide range of finite energies. 
In Fig.~\ref{figspinon2}, we explictly compute the energy 
and momentum spread of the contribution to spin-spin correlation 
function as measured by neutron scattering due to the spinon 
particle-hole pair excitation  for the specific choice of spinon 
hoppings. Qualitatively a broad continuum is observed, 
with a small amount of missing weight near the  $\Gamma$ point 
due to features of the inter-band transition.

\begin{figure}[t]
{\includegraphics[width=6cm]{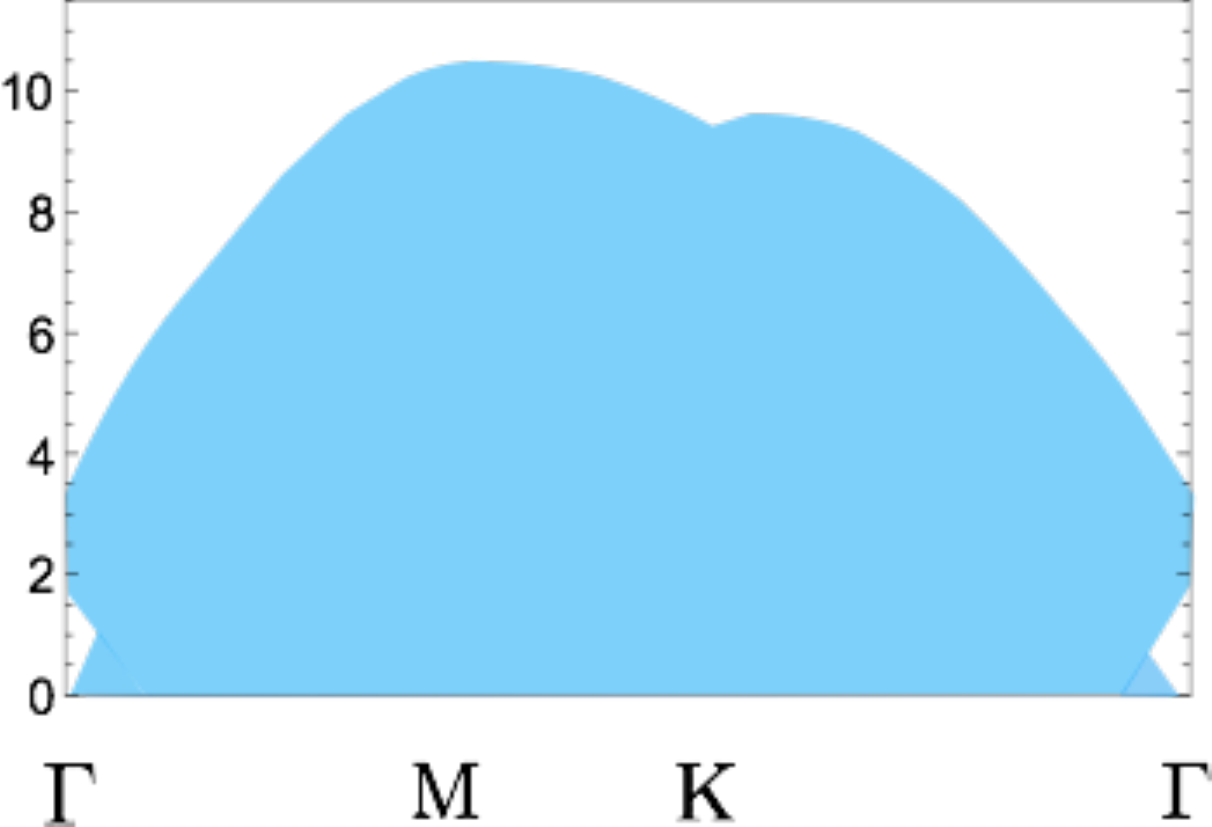}}
\caption{(Color online.) The spinon continuum contribution to the spin spin correlation 
function as measured by neutron scattering along high symmetry momentum direction. 
Due to the inter-band transition, there exists the spinon continuum up to finite 
energies near the $\Gamma$ point, with a small region of missing weight. 
The energy parameters here are the same as the ones in Fig.~\ref{figspinon}. }
\label{figspinon2}
\end{figure}

\section{Emergent orbital order in a field}
\label{sec6}

Despite the possible exotic spin liquid ground state at zero field, 
the Kugel-Khomskii model $H_{\text{KK}}$ becomes more tractable 
in the presence of a strong external magnetic field. Due to the 
suppression of the exchange coupling in $H_{\text{KK}}$, it is 
feasible to choose the magnetic fields to fully polarize the local 
spin moments such that ${s^z = \,\uparrow}$ for every resonating hexagon, 
but at the same time keep the field from 
polarizing all the electron spins in the kagome system. 
The pseudospin $\boldsymbol{\tau}$ is not directly effected by the 
magnetic field since it does not couple to the Zeeman field. 
The pseudospins remain active, and the interaction between them 
turns out to be a ferromagnetic compass model on the triangular 
lattice formed by the resonating hexagons, 
\begin{equation}
H_{\text{RKK}} = -\frac{2J'}{9}\sum_{\boldsymbol R} \sum_{\mu = x,y,z} 
T^\mu_{\boldsymbol R} T^\mu_{{\boldsymbol R} + {\boldsymbol a}_{\mu}}.
\end{equation}

From a standard Luttinger-Tisza type of mean-field approach~\cite{PhysRev.70.954}, 
we find that the mean-field ground state of $H_{\text{RKK}}$ 
has an accidental U(1) continuous degeneracy, {\it i.e.} any 
ferro-orbital (${{\boldsymbol q}={\boldsymbol 0}}$) state with the pseudospin 
$\boldsymbol{\tau}$ orienting in $xz$ plane is a classical ground state. 
Here, we parametrize the mean-field pseudospin order as 
\begin{equation}
\boldsymbol{\tau}_{\text{cl}} = \frac{1}{2} ( \cos \theta\, \hat{z} 
+ \sin \theta \, \hat{x})
\label{eq10}
\end{equation}
with ${\theta \in [0,2\pi) }$.

\begin{figure}[b]
{\includegraphics[width=6cm]{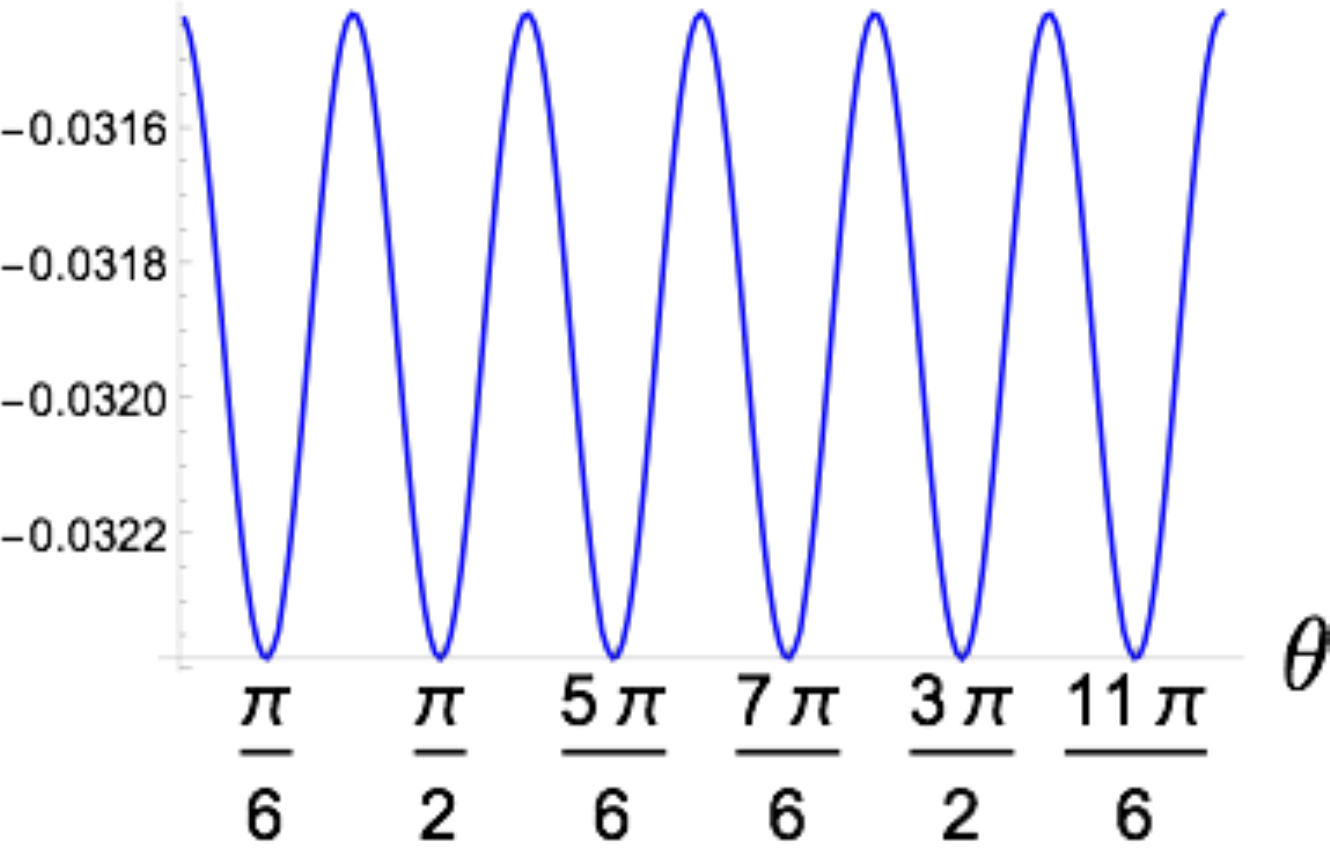}}
\caption{ 
The quantum zero-point energy per resonating hexagon for the 
mean-field orbital order. The energy unit is set to $2J'/9$ 
in the figure.
}
\label{fig5}
\end{figure}

This continuous U(1) ground state degeneracy at the mean-field level 
of the reduced Kugel-Khomskii model $H_{\text{RKK}}$ 
is lifted when the quantum fluctuations of the orbitals are included. 
We study this quantum order by disorder phenomenon from 
the linear orbital-wave theory. Here we introduce the Holstein-Primakoff
boson to represent the pseudospin operator $\boldsymbol{\tau}_{\boldsymbol R}$ 
as follows,
\begin{eqnarray}
&&\boldsymbol{\tau}_{\boldsymbol R}^{} \cdot \hat{\boldsymbol{\tau}}_{\text{cl}}^{} 
= \frac{1}{2} - a^\dagger_{\boldsymbol R} a_{\boldsymbol R}^{\phantom\dagger},
\\
&& \boldsymbol{\tau}_{\boldsymbol R}^{} \cdot  \hat{y} 
= \frac{1}{2i} [ a_{\boldsymbol R}^{\phantom\dagger} 
- a^\dagger_{\boldsymbol R} ],
\\
&&\boldsymbol{\tau}_{\boldsymbol R}^{} \cdot ( \hat{y} \times  
\hat{\boldsymbol{\tau}}_{\text{cl}}^{} )
= \frac{1}{2} [ a_{\boldsymbol R}^{\phantom\dagger} 
+ a^\dagger_{\boldsymbol R} ],
\end{eqnarray}
where ${\hat{\boldsymbol{\tau}}_{\text{cl}} \equiv 
{\boldsymbol{\tau}}_{\text{cl}}/|{\boldsymbol{\tau}}_{\text{cl}}|}$ 
is the orientation of the pseudospin.
We keep the quadratic terms in the Holstein-Primakoff boson operators and express
the reduced Kugel-Khomskii model as
\begin{eqnarray}
H_{\text{RKK}} &=& \sum_{{\boldsymbol k} 
\in \text{BZ}} \big[
2 A_{\boldsymbol k} a_{\boldsymbol k}^\dagger 
a_{\boldsymbol k}^{\phantom\dagger} 
+ B_{\boldsymbol k} (a_{\boldsymbol k}^{\phantom\dagger} 
a_{-{\boldsymbol k}}^{\phantom\dagger} 
+ h.c.)
\big]
\nonumber \\
&& \quad\quad\quad + E_{\text{cl}} ,
\end{eqnarray}
where ``BZ'' refers to the Brioullin zone of the triangular lattice
formed by the resonating hexagon plaquettes and 
\begin{widetext}
\begin{eqnarray}
E_{\text{cl}} &=& -\frac{J'}{12} \frac{N}{3} ,
\\
A_{\boldsymbol k} &=&  \frac{2J'}{9}\big[{ - \frac{\sin^2 ( \theta - \pi/3)}{4} 
\cos ({\boldsymbol k}\cdot {\boldsymbol a}_x)
 -  \frac{\sin^2 ( \theta +\pi/3)}{4} 
\cos ({\boldsymbol k}\cdot {\boldsymbol a}_y)
 - \frac{\sin^2 \theta}{4} \cos ({\boldsymbol k} \cdot {\boldsymbol a}_z) 
 +  \frac{3}{4} } \big],
\\
B_{\boldsymbol k} &=&  \frac{2J'}{9} \big[ {- \frac{\sin^2 ( \theta - \pi/3)}{4}  
\cos ({\boldsymbol k} \cdot {\boldsymbol a}_x)
 - \frac{\sin^2 ( \theta + \pi/3)}{4}  \cos ({\boldsymbol k} \cdot {\boldsymbol a}_y)-
 \frac{\sin^2 \theta}{4} \cos ({\boldsymbol k} \cdot {\boldsymbol a}_z)} \big]  ,
\end{eqnarray}
\end{widetext}

The linear orbital-wave Hamiltonian is then diagonalized by a 
Bogoliubov transformation for the Holstein-Primakoff bosons and 
is given by 
\begin{eqnarray}
H_{\text{RKK}} &=& E_{\text{cl}} +\sum_{{\boldsymbol k} \in \text{BZ}}
[\frac{\omega_{\boldsymbol  k}}{2}  -  A_{\boldsymbol k} ] \nonumber 
  + \sum_{{\boldsymbol k} \in \text{BZ}}
\omega_{\boldsymbol k}^{} \alpha^\dagger_{\boldsymbol k} 
\alpha_{\boldsymbol k}^{\phantom\dagger} ,
\label{eq11}
\end{eqnarray}
where the orbital-wave (or `orbiton') mode reads 
\begin{equation}
\omega_{\boldsymbol k} = 2 {(A_{\boldsymbol k}^2 - B_{\boldsymbol k}^2)^{ \frac{1}{2}  }}.
\end{equation}
From Eq.~\eqref{eq11}, the quantum correction to the ground state energy is 
\begin{equation}
\Delta E = \sum_{{\boldsymbol k} \in \text{BZ}} 
[\frac{\omega_{\boldsymbol k}}{2} - A_{\boldsymbol k} ]. 
\end{equation}
In Fig.~\ref{fig5}, we plot the quantum correction as a function of the 
angular parameter $\theta$. The minima occur at 
\begin{eqnarray}
{\theta = \frac{\pi}{6}+ \frac{ n\pi}{3}} ,  \quad {n \in {\mathcal Z}}, 
\end{eqnarray}
and are indicated in Fig.~\ref{fig6}a.

\begin{figure}[b]
{\includegraphics[width=6cm]{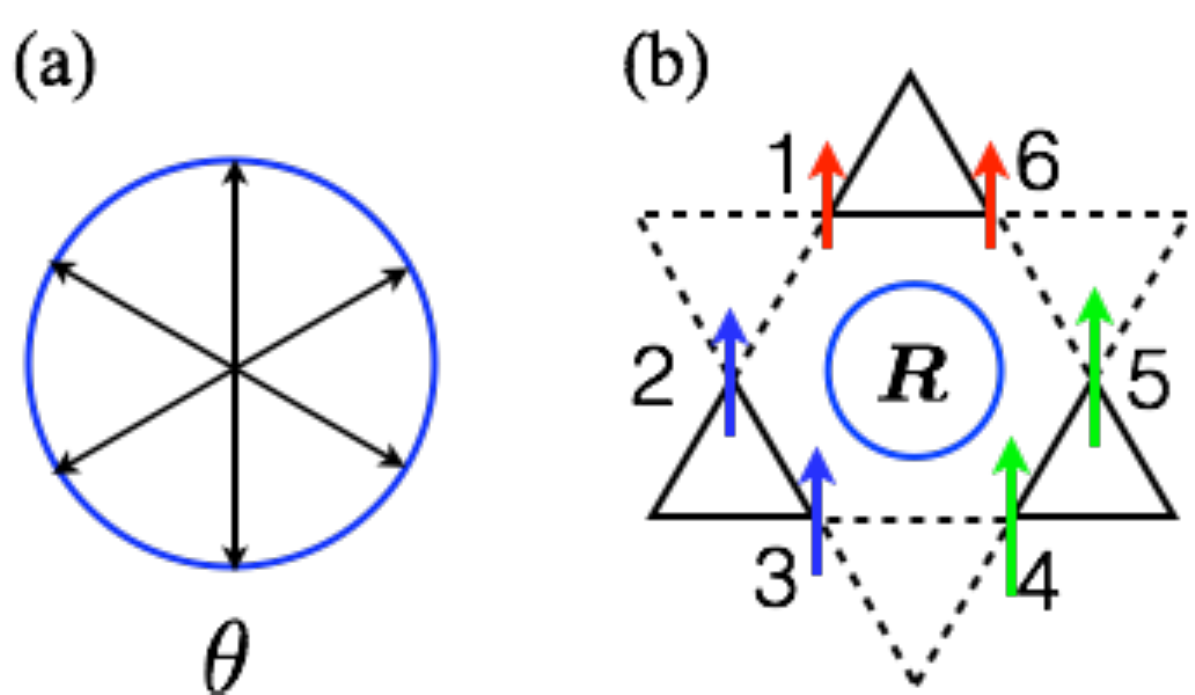}}
\caption{(Color online.) (a) The selection of the $\theta$ on a unit circle
by quantum fluctuation. The arrow indicates the optimal chioce. 
(b) The magnetic moment distribution within the resonating hexagon 
for $\theta = \pi/6$. It is clear that the three-fold rotation about 
the center of the hexagon is broken.}
\label{fig6}
\end{figure}

Since the two-fold orbital degeneracy arises from the point group symmetry, 
the emergent orbital order, that breaks the orbital degeneracy, has to be 
related to the symmetry breaking. To understand the physical consequence
of the orbital order, we consider the following product state wavefunction 
that is appropriate for the ${{\boldsymbol q} = {\boldsymbol 0}}$ 
ferro-orbital state,
\begin{equation}
|\Psi\rangle_{\text{orb}}^{} = \prod_{\boldsymbol R} \big[ 
\cos \frac{\theta}{2} 
| {\uparrow \uparrow} \rangle_{\boldsymbol R}^{} 
+ \sin \frac{\theta}{2} 
| {\downarrow \uparrow} \rangle_{\boldsymbol R}^{} \big],
\end{equation}
This variational wavefunction gives the orbital ordering 
in Eq.~\eqref{eq10}. From this wavefunction, we find that the electron 
density is uniform at every site within each resonating hexagon
and thus preserves the rotation and reflection symmetries. 
We then compute the local magnetization for each site within the 
resonating hexagon, 
\begin{eqnarray}
\langle s_1^z\rangle_{\boldsymbol R}^{} = \langle s_6^z \rangle_{\boldsymbol R}^{} &=& 
\frac{1}{12}+ \frac{\sin (\theta-{\pi}/{6})}{6} ,
 \\
\langle s_2^z\rangle_{\boldsymbol R}^{} = \langle s_3^z \rangle_{\boldsymbol R}^{} &=& 
\frac{1}{12} + \frac{\sin \theta}{6},
 \\
\langle s_4^z\rangle_{\boldsymbol R}^{} = \langle s_5^z \rangle_{\boldsymbol R}^{} &=& 
\frac{1}{12} - \frac{\sin( \theta + {\pi}/{6})}{6}.
\end{eqnarray}
Although the total local magnetization of each resonating hexagon 
is ${\langle s^z \rangle_{\boldsymbol R}^{}
= \sum_{i=1}^6 \langle s^z_i \rangle_{\boldsymbol R}= 1/2}$, 
the orbital ordering leads to a modulation of the spin ordering 
inside each resonating hexagon (see Fig.~\ref{fig6}b). 
The three-fold rotational symmetry about the 
center of the resonating hexagon is explicitly broken
by the orbital ordering. 

\begin{figure}[t]
{\includegraphics[width=6cm]{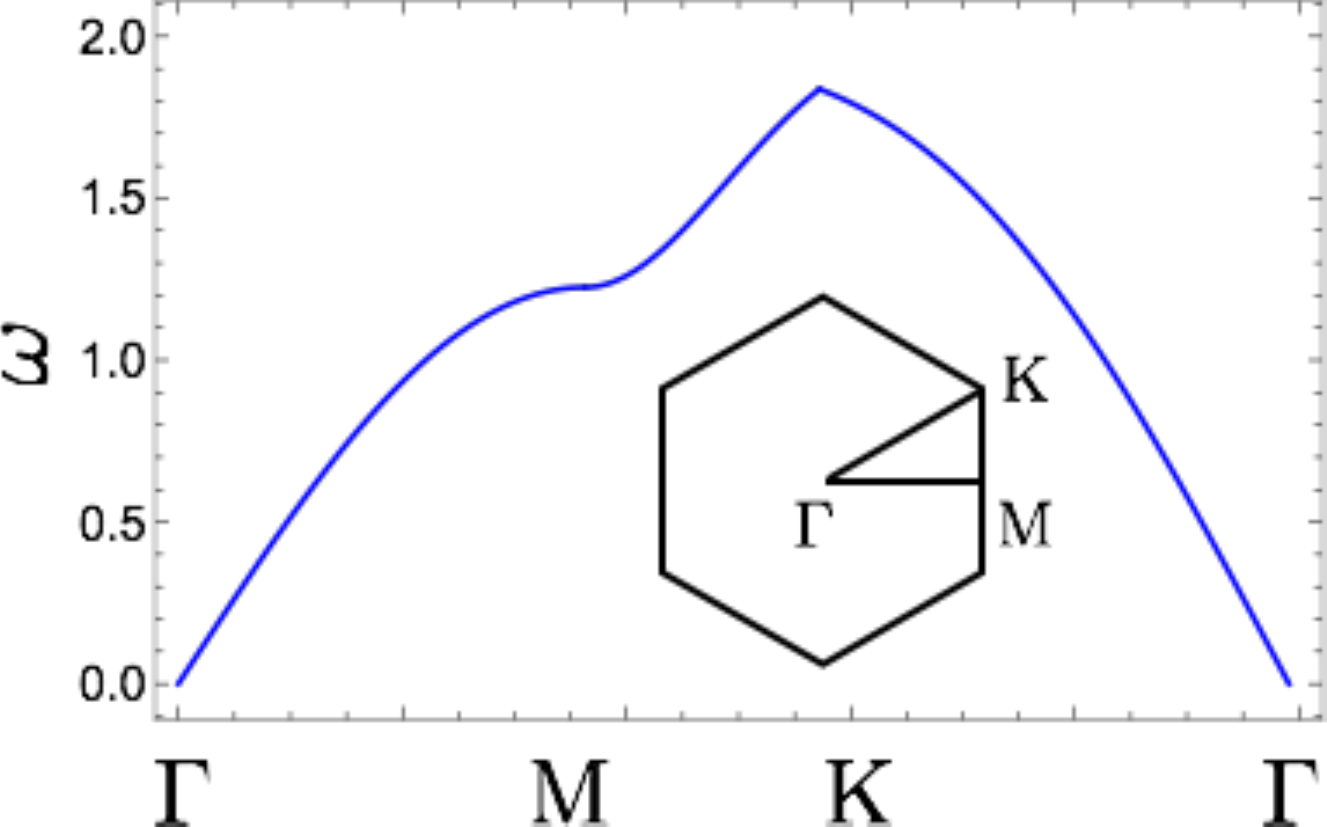}}
\caption{The orbiton disperson along the high-symmetry momentum line.   
The inset is the Brioullin zone of the triangular lattice formed 
by the resonating hexagons. The energy unit is set to $2J'/9$ in 
the figure.
}
\label{fig7}
\end{figure}

In Fig.~\ref{fig7}, we plot the dispersion of the orbiton excitation for 
$\theta=\pi/6$. We find the dispersion is gapless at the $\Gamma$ point 
due to the breaking of the accidental U(1) degeneracy. This pseudo-Goldstone 
mode is expected to be gapped if the interaction between the Holstein-Primakoff 
bosons is included. Since the interaction induced gap should be 
very small compared to the orbiton energy scale, one would expect 
to observe the heat capacity $C_v \sim T^2$ at low temperatures. 

\section{Discussion}
\label{sec7}


We discuss the experimental consequences of the plaquette charge order, 
the emergent orbitals, and the orbital orders.
The plaquette charge order explicitly breaks the lattice translation symmetry
and would lead to some variation of the bond lengths according to the symmetry
breaking. This may be detected by high-resolution X-ray scattering or X-ray
pair distribution function (PDF) measurement. The plaquette charge order 
reconstructs the spin states within each resonant hexagon 
leading to the freezing of 2/3 of the spins, 
as observed in the spin susceptibility in 
LiZn$_2$Mo$_3$O$_8$~\cite{Mourigal14,Sheckelton14,Sheckelton12,Mcqueen2015,Quilliam}. 
A different explanation of the susceptibility anomaly in LiZn$_2$Mo$_3$O$_8$ 
based on the lattice distortion and the emergent lattices has been proposed  
in a previous work~\cite{Flint13}. Both this previous work and the current 
work require a translation symmetry breaking by tripling the crystal unit 
cell. Such a translation symmetry breaking has not yet been observed in 
the experiment.  Here we point out the possible reason, namely, 
that under certain conditions, the symmetry breaking must be short 
range at all temperatures. 

\begin{figure}[t]
{\includegraphics[width=5cm]{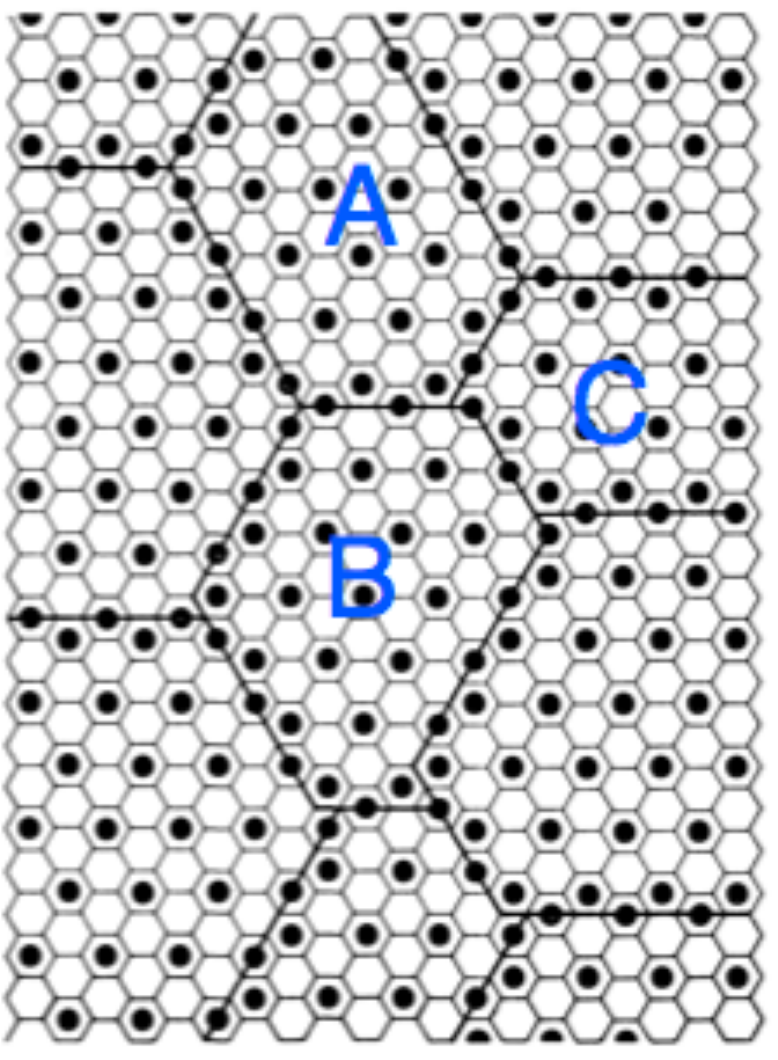}}
\caption{A picture of the domain walls separating the ABC domains 
when the electron occupation is off-stoichiometry, 
in this case slightly more than 1/6 per Mo.
}
\label{fig11}
\end{figure}

The Li ion is mobile and may make the system slightly off stoichiometry. 
To accommodate the missing or extra charges, the system needs to create 
domain walls within the symmetry broken phase. An example of such domain 
walls is shown in Fig.~\ref{fig11} for the case when the filling is slightly more 
than 1/6. Each solid dot represents the charge order shown in Fig.~\ref{fig1}. 
Note that the charge order can be centered on one of three hexagons, 
thereby forming ABC type domains. A certain density of domain walls 
will be required for a given deviation from from 1/6 filling. 
There is an energy cost per unit length of the domain wall, 
because electrons are now forced to occupy neighboring sites. 
The resulting state is expected to be a ``liquid'' 
state with an exponential decay of correlations for the electron charge 
density~\cite{PhysRevB.25.349}. This result is special for 
the hexagonal arrangement of domain walls and the reason is as follows.  
It was pointed out by J.~Villain~\cite{Villain}, there exists a 
breathing mode that expands or shrinks one particular domain but 
costs no energy because the total domain wall length is not changed. 
This is illustrated in Fig.~\ref{fig12}. Consequently the free energy of the 
system of domain walls is purely entropic and is proportional to 
temperature $T$. In the long wavelength limit, the elastic constant 
of the system is also proportional to $T$ and so is the energy to 
create a dislocation. An example of a dislocation is shown in Fig.~\ref{fig13}. 
By the usual Berezinski-Kosterlitz-Thouless (BKT) argument, 
the competition of this energy with the entropy associated with 
the dislocation determines whether the dislocation will proliferate, 
resulting in an expontially decaying correlation function. 
Unlike the usual BKT argument, where the dislocation enery 
is a constant and a phase transition is predicted at a finite 
temperature, here the result depends on the numerial coefficient 
of the linear $T$ term in the elastic energy. A detailed computation 
carried out in Ref.~\onlinecite{PhysRevB.25.349} showed that the 
system is always disordered at any temperature. A short range 
charge ordering  makes the detection more difficult, but not 
impossible. Perhaps resonant X-ray scattering which couples 
directly to the electrons will have a better chance seeing this distortion.

\begin{figure}[t]
{\includegraphics[width=4.5cm]{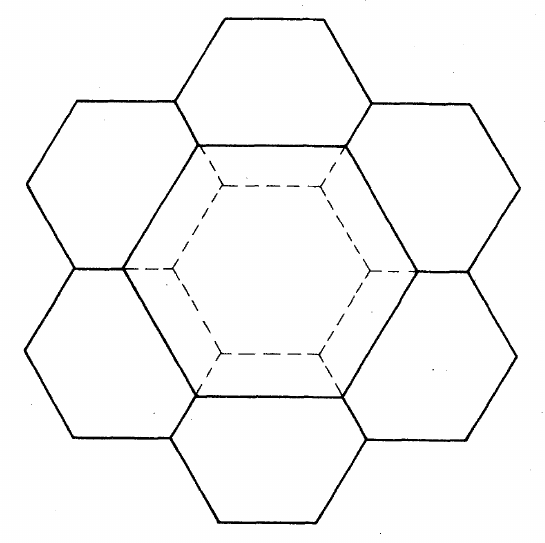}}
\caption{The breathing mode of Villain. 
Note that the total wall length and hence 
its energy has not changed. This mode contribute only to the entropy.
}
\label{fig12}
\end{figure}

The emergent orbital is a degree of freedom that naturally emerges from the 
plaquette charge order on the breathing Kagome lattice. The presence of this extra 
degree of freedom distinguishes the current proposal from the previous one
in Ref.~\onlinecite{Flint13}. However, the emergent orbital is not detectable 
in the magnetization measurement since the orbital does not couple directly to 
the external magnetic field. However, 
it does contribute to the heat capacity and the entropy. We expect an additional 
entropy from the emergent orbitals apart from the spin entropy. This extra 
entropy from the orbitals may be measurable on the high-quality samples. 
The suggested spinon Fermi surface ground state and the spinon excitation
should be detectable via inelastic neutron scattering. In fact, the existing 
measurement does suggest a broad continuum of excitations~\cite{Mourigal14},
even though the measurement was taken on powder samples. 
Since the qualitative feature for the spinon inter-band particle-hole excitation
is more visible at the $\Gamma$ point,  optical measurement or Raman scattering 
can be useful for detecting the finite energy spinon continuum at the $\Gamma$ point. 

A magnetic field that is of the order of the low-temperature
Curie-Weiss temperature is expected to polarize the spin degree of freedom.
The magnetic field should be much less than the high temperature Curie-Weiss
temperature to prevent polarizing the spins that form the spin singlet within
the resonating hexagon. The remaining orbital degrees of freedom then 
develop an orbital order via a quantum order by disorder mechanism
and support a pseudo-Goldstone mode that gives a heat capacity 
${C_v \sim T^2}$ at low temperatures. The orbital wave excitation 
may be detected by the resonant inelastic X-ray scattering.  
The orbital order creates a magnetic moment redistribution within the 
resonating hexagon. This intra-hexagon static magnetic structure may be 
detected by high-resolution neutron scattering, $\mu$SR and/or NMR measurements.

\begin{figure}[htp]
{\includegraphics[width=7.5cm]{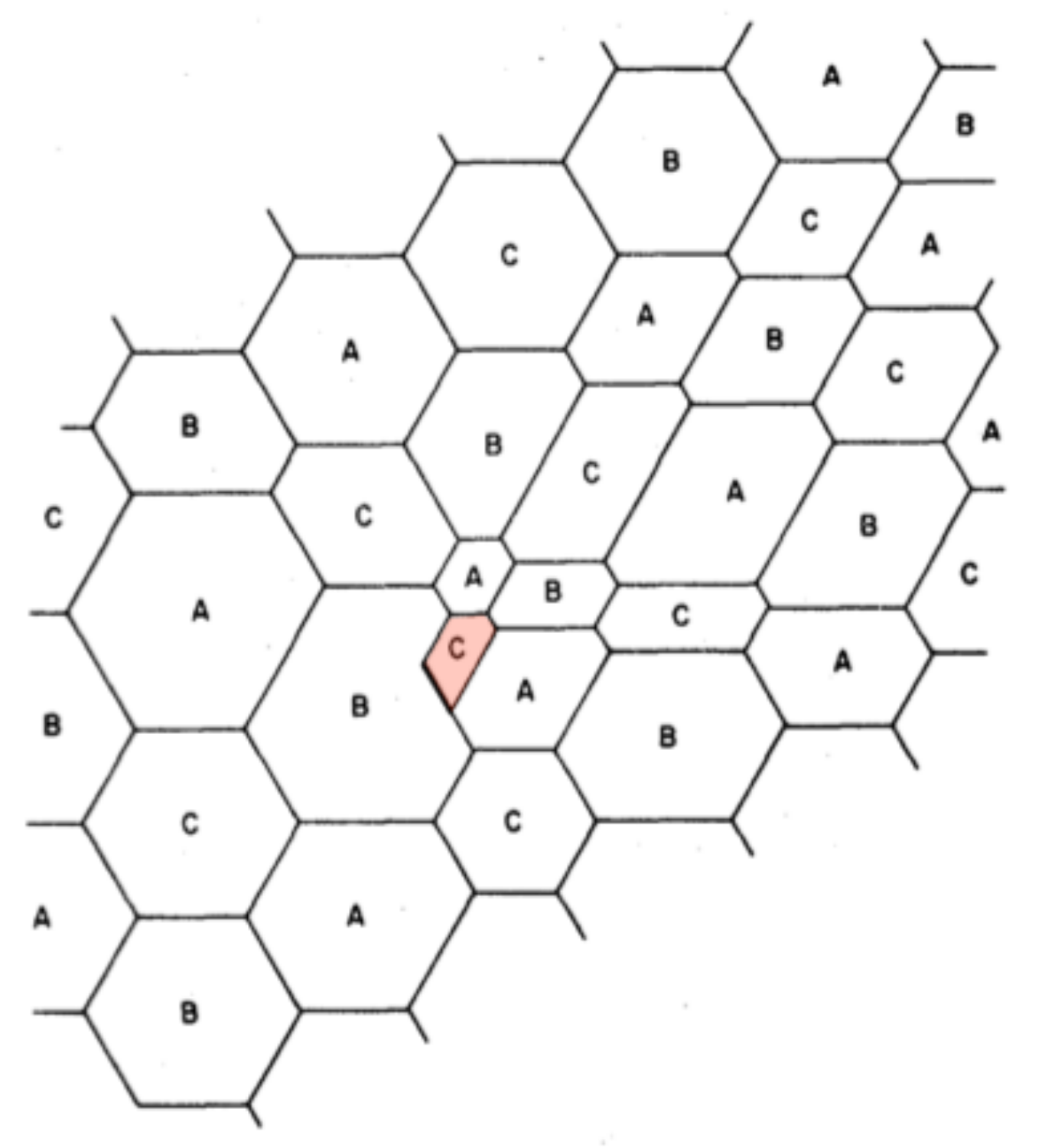}}
\caption{A picture of a dislocation center (in red) in the system of domain walls. 
}
\label{fig13}
\end{figure}

Finally, there exists a large family of cluster magnets in which the electrons
are localized on the cluster units and form CMIs~\cite{ChenPRL2014,ChenPRB2016,Lv2015,PhysRevB.96.054405,
KimHS2014,PALeeLaw,PhysRevLett.93.126403,PhysRevLett.110.166402,CheongNature,GALL201399,doi:10.1021/ic00198a009}. 
The physical properties of many of these cluster magnets have   
not been explored carefully. Recently, 1T-TaS$_2$ is proposed 
as a QSL candidate~\cite{PALeeLaw}. In this system, the low-temperature
(commensurate) charge density wave order enlarges the unit cell 
such that there exists one localized and unpaired electron inside 
the 13-site David-star cluster. This system can thus be considered 
as a CMI~\cite{PALeeLaw}. These clusterly localized electrons 
form effective spin-1/2 local moments that interact with each 
other and may develop a spin liquid ground state~\cite{PALeeLaw}. 
Besides these two-dimensional cluster magnets, the Ta-based and 
Mo-based lacunar spinels are good examples of three dimensional 
CMIs~\cite{ChenPRL2014,KimHS2014,PhysRevLett.93.126403,
PhysRevLett.110.166402,PhysRevLett.93.126403}. 
In these materials, the systems naturally host a breathing
pyrochlore lattice structure where one half of the tetrahedral 
clusters is smaller than the other half and host the localized 
electrons~\cite{ChenPRL2014}. The study on these systems are 
quite limited so far. We expect that the cluster localization 
of the electrons in these CMIs may bring some interesting 
phenomena and enriches our understanding of Mott physics. 

\section{Acknowledgments}

G.C. is supported by the ministry of science and technology of 
China with the grant No.2016YFA0301001, the reserach initiative funds 
and the program of first-class university construction of Fudan University, 
and the thousand-youth-talent program of China. P.A.L. is supported 
by the U. S. National Science Foundation under DMR-1522575.

\bibliography{ref}

\end{document}